\documentclass[letterpaper,twocolumn,10pt]{article}
\usepackage{usenix-2020-09}

\usepackage{tikz}
\usepackage{amsmath}

\newif\ifanon
\anonfalse

\newif\ifdraft
\draftfalse

\usepackage{adjustbox}
\usepackage{hyperref}
\usepackage[ruled,vlined,linesnumbered]{algorithm2e} %
\usepackage{amssymb}
\usepackage[capitalise,noabbrev]{cleveref}
\usepackage{siunitx}
\usepackage[ff,keys,primitives,operators]{cryptocode}
\usepackage{subcaption}
\usepackage{marvosym}
\usepackage{acro}
\usepackage{booktabs}
\usepackage{pgfplots}
\pgfplotsset{compat=1.18}
\usepackage{multirow}
\usepackage{multicol}
\usepackage{makecell}
\usepackage{comment}
\usepackage{blindtext}
\usepackage{tablefootnote}
\usepackage{pifont}
\usepackage{subcaption}

\usepackage{orcidlink}

\usetikzlibrary{arrows, automata, positioning}
\usetikzlibrary{shapes.geometric, arrows}
\usetikzlibrary{shadings}

\usepackage{todonotes}
\pgfplotsset{
  hammer time series/.style={
    xmin=0, xmax=20,
    ymin=0,
    axis background/.style={%
        preaction={
            path picture={
                \foreach \row/\col/\label in \Rows {
                    \edef\temp{
                        \noexpand\draw[\col,fill=\col] (axis cs:0,\row) rectangle (axis cs:20,\row+1);
                    }
                    \temp
                }
            }
        }
    },
    clip=false,
    width=10cm,
    xlabel={Time},
    grid=major,
    ytick distance=1,
    yticklabels={},
    xtick={0,1,...,20},
    tick label style={font=\small},
    label style={font=\small},
  }
}

\usetikzlibrary{
  arrows,
  arrows.meta,
  automata,
  positioning,
  decorations.pathreplacing,
  shapes.geometric,
}

\DeclareAcronym{RSA}{
    short = RSA,
    long  = Rivest-Shamir-Adleman Algorithm,
}

\DeclareAcronym{DSA}{
    short = DSA,
    long  = Digital Signature Algorithm,
}

\DeclareAcronym{ECDSA}{
    short = ECDSA,
    long  = Elliptic Curve Digital Signature Algorithm,
}

\DeclareAcronym{WOTSP}{
    short = WOTS\texorpdfstring{\textsuperscript{+}}{+},
    long  = Winternitz One-Time Signature Scheme\texorpdfstring{\textsuperscript{+}}{+},
}
\newcommand{\WOTSP}{\acthe{WOTSP}}

\DeclareAcronym{XMSS}{
    short = XMSS,
    long  = eXtended Merkle Signature Scheme,
}
\newcommand{\XMSS}{\acthe{XMSS}}

\DeclareAcronym{PQC}{
    short = PQC,
    long  = Post-Quantum Cryptography,
}
\newcommand{\PQC}{\ac{PQC}}

\DeclareAcronym{FORS}{
    short = FORS,
    long  = Forest of Random Subsets,
}
\newcommand{\FORS}{\ac{FORS}}

\DeclareAcronym{SLH-DSA}{
    short = SLH-DSA,
    long  = Stateless Hash-Based Digital Signature Algorithm,
}
\newcommand{\SLHDSA}{\ac{SLH-DSA}}

\DeclareAcronym{RFM}{
    short=RFM,
    long = Refresh Management,
}

\DeclareAcronym{PRAC}{
    short=PRAC,
    long=Per Row Activation Counting
}

\DeclareAcronym{TRR}{
    short = TRR,
    long  = Target Row Refresh,
}
\newcommand{\TRR}{\ac{TRR}}

\DeclareAcronym{DRAM}{
    short = DRAM,
    long  = Dynamic Random Access Memory,
}
\newcommand{\DRAM}{\ac{DRAM}}

\DeclareAcronym{SRAM}{
    short = SRAM,
    long  = Static Random Access Memory,
}

\DeclareAcronym{TPM}{
    short = TPM,
    long  = Trusted Platform Module,
}

\DeclareAcronym{NIST}{
    short = NIST,
    long  = National Institute of Standards and Technology,
}

\DeclareAcronym{PFN}{
    short = PFN,
    long  = Page Frame Number,
}

\DeclareAcronym{AREF}{
    short = AREF,
    long  = Auto-Refresh,
}

\DeclareAcronym{THP}{
    short = THP,
    long  = Transparent Huge Page,
}

\DeclareAcronym{MTRRs}{
    short = MTRRs,
    long  = Memory Type Range Registers,
}

\DeclareAcronym{NUMA}{
    short = NUMA,
    long  = Non-Uniform Memory Access,
}

\DeclareAcronym{ASLR}{
    short = ASLR,
    long  = Address Space Layout Randomization,
}

\DeclareAcronym{MMU}{
    short = MMU,
    long  = Memory Management Unit,
}
\newcommand{\MMU}{\ac{MMU}}

\DeclareAcronym{ECC}{
    short = ECC,
    long  = Error-Correcting Codes,   
}
\newcommand{\ECC}{\ac{ECC}}

\DeclareAcronym{TEE}{
    short = TEE,
    long  = Trusted Execution Environment,
}

\acsetup{format/long={\emph}}

\DeclareAcroArticle{definite}{the}
\newcommand{\acthe}[1]{\ac[first-style=long-short-the]{#1}}

\sfcode`\.=1001
\sfcode`\?=1001
\sfcode`\!=1001
\sfcode`\:=1001

\makeatletter
\newif\ifsmartthe@start
\smartthe@starttrue     %

\AddToHook{para/begin}{\global\smartthe@starttrue}

\newcommand{\smartthe}{%
  \ifsmartthe@start
    \global\smartthe@startfalse
    The
  \else
    \ifnum\spacefactor>1000
      The
    \else
      the
    \fi
  \fi
}
\makeatother

\NewAcroTemplate{long-short-the}{%
  \acroiffirstTF{%
    \acrowrite{long}(%
      \acroifT{foreign}{\acrowrite{foreign}, }%
      \acrowrite{short}%
      \acroifT{alt}{ \acrotranslate{or} \acrowrite{alt}}%
      \acrogroupcite
    )%
    \acspace
  }{%
    \acrowrite{short}%
  }%
}

\newcommand{\RomanNumber}[1]{\textup{\uppercase\expandafter{\romannumeral#1}}}

\newcommand{\instr}[1]{\texttt{#1}}
\newcommand{\var}[1]{\texttt{#1}}
\newcommand{\tech}[1]{\textsc{#1}}
\newcommand{\file}[1]{\texttt{#1}}
\newcommand{\func}[1]{\texttt{#1}}

\def\sig{\sigma}

\newcommand{\wsig}{\sig^W}

\newcommand{\ADRS}{\textbf{ADRS}}
\newcommand{\hatSKseed}{\hat{\textbf{SK}}\text{.seed}}
\newcommand{\SKseed}{\textbf{SK}\text{.seed}}
\newcommand{\PKseed}{\textbf{PK}\text{.seed}}
\newcommand{\PKroot}{\textbf{PK}\text{.root}}

\newcommand{\SPHINCSP}{SPHINCS\texorpdfstring{\textsuperscript{+}}{+}}

\newcommand{\lgw}{\lg_w}

\newcommand{\sfh}{\textsf{h}}
\newcommand{\auth}[1]{\mathop{auth}(#1)}

\newcommand{\byte}{\mathbb{B}}

\def\rowwidth{3cm}
\def\rowheight{1cm}

\definecolor{aggressor}{RGB}{238,153,170}       %
\definecolor{victim}{RGB}{102,153,204}          %
\definecolor{accesscol}{RGB}{238,102,119}
\definecolor{emptycol}{RGB}{187,187,187}

\definecolor{brightblue}{RGB}{0,68,136}
\definecolor{brightred}{RGB}{187,85,102}
\definecolor{brightyellow}{RGB}{221,170,51}
\definecolor{brightgreen}{RGB}{34,136,51}

\newcommand{\graftingred}{brightred!50!red}

\definecolor{lightblue}{RGB}{102,153,204}
\definecolor{lightred}{RGB}{238,153,170}
\definecolor{lightyellow}{RGB}{238,204,102}

\definecolor{mygreen}{RGB}{34,136,51}
\definecolor{mygrey}{RGB}{187,187,187}

\newcommand{\drawMemoryPattern}[1]{
    \foreach[count=\i from -3] \label/\col in {#1} {%
        \fill[\col] (0,-\i*\rowheight) rectangle (\rowwidth,-\i*\rowheight - \rowheight);%
        \draw (0,-\i*\rowheight) rectangle (\rowwidth,-\i*\rowheight - \rowheight);%
        \ifx\label\empty\else%
            \node at ({\rowwidth/2}, {-\i*\rowheight - \rowheight/2}) {\textbf{\label}};%
        \fi%
    }%
}

\pgfplotsset{
  compat=1.18,
}

\makeatletter
\DeclareRobustCommand{\rvdots}{%
  \vbox{
    \baselineskip4\p@\lineskiplimit\z@
    \kern-\p@
    \hbox{.}\hbox{.}\hbox{.}
  }}
\makeatother

\newcommand{\MiB}[1]{\qty{#1}{\mebi\byte}}
\newcommand{\GiB}[1]{\qty{#1}{\gibi\byte}}

\newcommand{\FnLink}[2]{\hyperref[#1]{%
  \textbf{%
    \color{black}{\func{#2}}%
  }}}%

\tikzset{
    dot/.style = {circle, fill, minimum size=#1,inner sep=0pt, outer sep=0pt},
    triangle/.style={isosceles triangle,draw,shape border rotate=90,isosceles triangle stretches=true, minimum width=15mm,inner sep=1mm},
    tree/.style = {triangle,draw,isosceles triangle apex angle=110}
}

\NewDocumentCommand{\wotssig}{O{$\sig_{#2}^W$} m m}{
{\begingroup
  \def\wsig{#1}
  \def\level{#2}
  \def\child{#3}
  \node[rectangle, draw, above=5mm of \child] (w\level) {WOTS\textsuperscript{+}};
  \node[right=1cm of w\level.east] (s\level) {\wsig};
  \draw[->,dotted] (w\level) -- (s\level);
\endgroup}}

\NewDocumentCommand{\xmss}{O{$\pk^X_{#3}$} O{$\auth{\pk_{#3}^W}$} m m}
{{\begingroup
  \def\root{#1}
  \def\xauth{#2}
  \def\level{#3}
  \def\child{#4}
  \node[tree,above=5mm of \child.north west, anchor=south east] (xmsstr\level) {XMSS};
  \node[dot=3pt,draw,fill=black] (xmssdot\level) at (xmsstr\level.north) {};
  \node[above=0.01cm of xmsstr\level.north] (xmss\level) {\root};
  \node[right=1cm of xmsstr\level.east] (xmssauth\level) {\xauth};
  \draw[->,dotted] (xmsstr\level) -- (xmssauth\level);
  \node[rectangle, draw, below=5mm of xmsstr\level.south west, anchor=north east] (wots\level1) {WOTS\textsuperscript{+}};
  \draw[->,dashed] (wots\level1.north) -- (xmsstr\level.south west -| wots\level1.north);
  \draw[->,dashed] (\child.north) -- (xmsstr\level.south east -| \child.north);
\endgroup}}

\ifdraft

\else
\excludecomment{draft}
\fi

\DeclareMathOperator{\toInt}{to\_int}

\date{}

\title{\Large \bf SLasH-DSA: Breaking SLH-DSA Using an Extensible End-To-End Rowhammer Framework}

\DeclareRobustCommand*{\authorrefmark}[1]{%
  \raisebox{0pt}[0pt][0pt]{\textsuperscript{\footnotesize #1}}%
}

\ifanon
\author{Anonymous Submission}
\else
\author{
    {\orcidlink{0009-0006-9189-3458} \rm Jeremy Boy}\\
    University of Luebeck
    \and
    {\orcidlink{0000-0002-1426-0853} \rm Antoon Purnal\authorrefmark{*}}\\
    Google
    \and
    {\orcidlink{0000-0001-7828-2333} \rm Anna P{\"a}tschke}\\
    University of Luebeck
    \and
    {\orcidlink{0009-0005-5785-9179} \rm Luca Wilke\authorrefmark{\dag}}\\
    Azure Research, Microsoft
    \and
    {\orcidlink{0000-0003-1116-6973} \rm Thomas Eisenbarth}\\
    University of Luebeck
}
\fi

\begin{document}

\maketitle

\ifanon
\else
\begingroup
  \renewcommand\thefootnote{\fnsymbol{footnote}}
  \footnotetext[1]{Work partially conducted while at PQShield.}
  \footnotetext[2]{Work partially conducted while at University of Luebeck.}
\endgroup
\fi

\begin{abstract}

As quantum computing advances, \PQC{} schemes are adopted to replace classical algorithms.
Among them is the \SLHDSA{} that was recently standardized by NIST and is favored for its conservative security basis. %

In this work, we present the first software-only universal forgery attack on \SLHDSA{}, leveraging Rowhammer-induced bit flips to corrupt the internal state and forge signatures.
While prior work targeted embedded systems and required physical access, our attack is software-only, targeting commodity desktop and server hardware, significantly broadening the threat model.
We demonstrate full end-to-end attacks against \SLHDSA{} in OpenSSL~3.5.1, achieving universal forgery for the SHAKE-128f (deterministic), SHA2-128s, and SHAKE-192f (randomized) parameter sets after one hour (deterministic) or eight hours (randomized) of hammering and post-processing ranging from minutes to an hour, and showing theoretical attack complexities for most parameter sets.
Our post-processing is informed by a novel complexity analysis that, given a concrete set of faulty signatures, identifies the most promising computational path to pursue.

To enable the attack, we introduce \tech{Swage}, a modular and extensible framework for implementing end-to-end Rowhammer-based fault attacks.
\tech{Swage} abstracts and automates key components of practical Rowhammer attacks.
Unlike prior tooling, \tech{Swage} is untangled from the attacked code, making it reusable and suitable for
frictionless analysis of different targets.
Our findings highlight that even theoretically sound \PQC{} schemes can fail under real-world conditions, underscoring the need for additional implementation hardening or hardware defenses against Rowhammer.

\end{abstract}

\section{Introduction}
\acreset{PQC}
\acreset{SLH-DSA}

With the accelerating progress in practical quantum computing, the cryptographic community is undergoing a major transition toward the adoption of \PQC{}.
In response, 
NIST has released several \PQC{} standards, including FIPS 205~\cite{nistStatelessHashBasedDigital2024} that describes the \SLHDSA{} signature scheme (known as \SPHINCSP{}~\cite{aumassonSPHINCSSubmissionNIST2022} prior to standardization).
However, the adoption of strong cryptographic primitives alone is not sufficient to achieve security in real-world deployments.
Besides being cryptographically sound, the implementation must also be able to withstand the various (micro)architectural side-channel and hardware-level attacks bedeviling modern CPUs.

One prominent and persistent threat is the \emph{Rowhammer} vulnerability.
Since its discovery in 2014~\cite{kimFlippingBitsMemory2014}, Rowhammer has evolved from a reliability issue into a powerful attack vector capable of inducing targeted bit flips in \DRAM{} through frequent accesses to adjacent memory rows.
Despite years of developing countermeasures, Rowhammer remains a viable threat even on the most recent DDR5 memory modules~\cite{cojocarExploitingCorrectingCodes2019, frigoTRRespassExploitingMany2020, jattkeBlacksmithScalableRowhammering2022, jattkeZenHammerRowhammerAttacks2024a}.
Over the past decade, it was shown how to leverage Rowhammer to tamper with isolation, escalate privileges, and leak secret data~\cite{seabornProjectZeroExploiting2015,amerPQHammerEndtoEndKey2025,kwongRAMBleedReadingBits2020,haidarCrowhammerFullKey2025,dioHalfSpectreFull2025,DBLP:conf/uss/WangPHSFK22}. %
Rowhammer-based attacks can also operate remotely via the browser or network requests~\cite{grussRowhammerjsRemoteSoftwareInduced2016,lippNethammerInducingRowhammer2020}.

As quantum-resilient schemes become part of the critical infrastructure, it becomes vital to understand their resilience against practical fault attacks.
In this work, we examine the vulnerability of the NIST-standardized post-quantum signature scheme \SLHDSA{} to Rowhammer-based fault attacks.
As a hash-based scheme, \SLHDSA{} derives its resistance against large-scale quantum computers solely from the security properties of hash functions, making it an attractive and conservative choice. 
Unlike other standardized hash-based signatures (XMSS and LMS)~\cite{cooper2020recommendation}, for which state-keeping is essential to maintain security, \SLHDSA{} is stateless, enabling it to act as a drop-in replacement for existing signature schemes.
While prior work demonstrated that physical fault injection can enable universal forgeries against \SLHDSA{}~\cite{genetProtectingSPHINCSFault2023}, we show for the first time that a similar attack is feasible entirely in software, requiring no physical access.
This significantly broadens the attack surface.

We demonstrate attacks against the \SLHDSA{} implementation in OpenSSL~3.5.1~\cite{opensslOpenSSL35Final2025}, showing practical forgery for the SHAKE-128f (deterministic), SHA2-128s and SHAKE-192f (randomized) parameter sets.
We show that an attacker can universally forge signatures for the \SLHDSA-SHAKE-192f parameter set after eight hours of Rowhammer attack and 1417 seconds of post-processing.
Besides practical forgery attacks, we show theoretical attack complexities for most parameter sets.
For \SLHDSA{}'s deterministic signing mode, which is not the default setting in OpenSSL but specified and allowed by the FIPS 205 standard, we achieve universal forgery for the \SLHDSA-SHAKE-128f parameter set after one hour of Rowhammer and 151 seconds of post-processing.

A given collection of faulty signatures leaves the attacker with several potential \SLHDSA{} subcomponents to target during the post-processing phase.
To select the target with the lowest expected runtime, we derive, for the first time, the offline attack complexity for a \emph{concrete} set of faulty signatures.

Performing end-to-end Rowhammer attacks in practice is non-trivial and entails several low-level challenges:
First, the attacker needs to reverse-engineer the mapping from physical addresses to \DRAM{} addresses for the CPU that they are using.
Based on that, they must find a hammering pattern that bypasses the \TRR{} mitigation on the used DIMM module.
Hammering patterns require a certain amount of physically contiguous memory.
However, unprivileged Linux users do not have an API that guarantees the allocation of such memory.
Thus, for a realistic attack, the adversary needs to resort to certain quirks of the Linux memory allocator and side-channels in the memory subsystem of the CPU.
Finally, the attacker needs to ensure that the victim application places the attackable data structure in a memory location they can manipulate through their hammering pattern.
While prior work~\cite{jattkeBlacksmithScalableRowhammering2022,adilettaMayhemTargetedCorruption2024,kwongRAMBleedReadingBits2020,DBLP:conf/uss/RazaviGBPGB16,bolcskeiRubiconPreciseMicroarchitectural2025} has addressed these steps, the provided artifacts either only solve parts of the attack chain or are closely coupled with the attacked application.

To provide a holistic solution, we introduce \tech{Swage}, a modular and extensible framework for implementing end-to-end Rowhammer-based fault attacks.
\tech{Swage} integrates modules for each of the mentioned attacks steps and can easily be applied to different victim applications.
In summary, our framework and the presented attack highlight the need to address Rowhammer and similar threats in the post-quantum era.

\subsection{Our Contributions}
\begin{itemize}
    \item We showcase a realistic Rowhammer attack against all parameter sets in OpenSSL's \SLHDSA\ implementation in both deterministic and randomized signing modes, including the end-to-end forgery of signatures for NIST security levels 1 and 3.
    \item We provide a detailed vulnerability analysis of \SLHDSA\ in the presence of Rowhammer-based faults.
    \item We derive the offline complexity given concrete faulty signatures, identifying the most promising forgery target.
    \item We implement \tech{Swage}, a modular Rowhammer attack framework to ease the implementation of Rowhammer attacks. We make our source code available on Github
    \ifanon
        \footnote{\href{https://anonymous.4open.science/r/SLasH-DSA}{https://anonymous.4open.science/r/SLasH-DSA}}.
    \else
        \footnote{\href{https://github.com/UzL-ITS/SLasH-DSA}{https://github.com/UzL-ITS/SLasH-DSA}}.
    \fi
 \end{itemize}

\subsection{Responsible Disclosure}

We officially reported our findings to the OpenSSL security team on August 04, 2025.
Our report was acknowledged on August 07, 2025.
We were informed that fault attacks are outside of OpenSSL's threat model, and permission was granted to release our findings.

\section{Background}

This section provides the technical context for understanding our Rowhammer-based attack on the \SLHDSA{} scheme.
We begin by describing the relevant characteristics of modern \DRAM{} enabling Rowhammer attacks.
We then summarize the structure and operation of the \SLHDSA{} signature scheme, highlighting components relevant to our attack strategy.

\subsection{Memory Management}

\acreset{DRAM}
Physical memory is implemented as an array of \DRAM\ cells, arranged in rows and columns within banks, which are in turn grouped into ranks.
Each bank includes a row buffer that contains the value of the most recently accessed row to speed up repeated read operations.
A \DRAM\ cell consists of a capacitor and a transistor; the capacitor's charge stores the bit's value, and the transistor provides access to the stored value.
As the charge leaks over time, the \DRAM\ cells must periodically be refreshed.

Modern operating systems (OS) use virtual memory to manage physical memory through a layer of indirection.
Processes work with virtual addresses that are then translated to physical addresses be the \MMU\ with the help of page tables.
After the translation of the virtual address to a physical address, the physical address is mapped to a physical location in \DRAM.
This mapping is CPU-specific and proprietary but can be reverse-engineered~\cite{pesslDRAMAExploitingDRAM2016,xiaoOneBitFlips2016}.

When a process requests memory, the OS needs to select the physical pages to back it. Linux relies on the buddy allocator strategy, which manages contiguous physical memory in buckets of increasing size. These can be split for smaller allocations and merged again when memory gets freed~\cite{bolcskeiRubiconPreciseMicroarchitectural2025}.

\subsection{Rowhammer}\label{background:rowhammer}%
\acreset{TRR}%
Rowhammer, discovered by Kim et al.~\cite{kimFlippingBitsMemory2014}, is a hardware vulnerability in \DRAM\ that allows an attacker to flip bits in inaccessible \emph{victim rows} by repeatedly accessing a set of neighboring \emph{aggressor rows} that are under their control.
After having mostly been considered a reliability issue at first, Rowhammer has been used to attack cryptographic schemes and circumvent OS isolation~\cite{seabornProjectZeroExploiting2015,amerPQHammerEndtoEndKey2025,kwongRAMBleedReadingBits2020,haidarCrowhammerFullKey2025,dioHalfSpectreFull2025}.
Manufacturers implemented various countermeasures such as \TRR\ in DDR4 memory, which identifies unusual memory access patterns and then issues refresh commands for suspected \emph{victim rows}.
However, research has shown that more complex hammering patterns can overload the access tracking in \TRR, re-enabling Rowhammer attacks~\cite{frigoTRRespassExploitingMany2020,jattkeBlacksmithScalableRowhammering2022}.
While the DDR5 standard has more principled mitigation~\cite{DBLP:journals/corr/abs-2406-19094,jedec2024ddr5c}, recent work~\cite{jattkeZenHammerRowhammerAttacks2024a} reports to have found at least one vulnerable setup.

\subsection{\SLHDSA}\label{background:slhdsa}
\SLHDSA\ is a hash-based digital signature scheme standardized by NIST in 2024~\cite{nistStatelessHashBasedDigital2024}. 
The building blocks of \SLHDSA\ are again hash-based signature schemes: (1) the few-time signature scheme \FORS, and (2) a hierarchical organization of the many-time signature scheme \ac{XMSS}.
The \XMSS\ trees, in turn, each manage a number of hash-based one-time signature scheme \ac{WOTSP} instances.
Although \FORS{} is an essential component of \SLHDSA{}, its specification is not relevant to understand the attack described in this paper.
For clarity, we omit its specification and refer to the FIPS 205 standard~\cite{nistStatelessHashBasedDigital2024}.

\NewDocumentCommand{\wotschain}{O{999} O{green} O{above} O{123} m}{{
\def\n{#1}
\def\c{#2}
\def\o{#3}
\def\l{#4}
\def\y{#5}
\node[square,fill={brightred}] (sk\y) at (0,\y) {\color{white!85!brightred} \small $\sk_{\l}$};
\node[round, fill={\ifnum\n=0 \c \else none\fi}] (i0\y) at (1.2,\y) {};
\node[round, fill={\ifnum\n=1 \c \else none\fi}] (i1\y) at (2.4,\y) {};
\node[round, fill={\ifnum\n=2 \c \else none\fi}] (i2\y) at (3.6,\y) {};
\node[round, fill={\ifnum\n=3 \c \else none\fi}] (i3\y) at (4.8,\y) {};
\draw[->] (sk\y) -- node[\o] {\small $\hash$} (i0\y);
\draw[->] (i0\y) -- node[\o] {\small $\hash$} (i1\y);
\draw[->] (i1\y) -- node[\o] {\small $\hash$} (i2\y);
\draw[->] (i2\y) -- node[\o] {\small $\hash$} (i3\y);
\pgfmathabs{\y-2.5}
\draw[->] (i3\y) -- ($(K.west)+(-.009*\pgfmathresult, \y*0.3-0.75)$);
}}

\NewDocumentCommand{\wotsgraph}{O{left} O{2} O{0} O{1} O{3} O{1} O{2}}{%
\tikzstyle{square} = [draw, minimum size=0.3cm]
\tikzstyle{round} = [circle, draw, minimum size=0.3cm]
\tikzstyle{process} = [rectangle, draw, minimum width=0.7cm, minimum height=1.9cm]
\node[process] (K) at (7,2.5) {\large $\fash$};
\node[square, fill=brightyellow] (out) at (8.5,2.5) {\small $\pk$};
\draw[->] (K) -- (out);
\wotschain[#2][brightgreen][#1][0]{5}
\wotschain[#3][brightgreen][#1][1]{4}
\wotschain[#4][brightgreen][#1][2]{3}
\wotschain[#5][brightgreen][#1][3]{2}
\draw[dashed] (-0.25,1.5) -- (5,1.5);
\wotschain[#6][brightgreen][#1][4]{1}
\wotschain[#7][brightgreen][#1][5]{0}
}

\begin{figure}
    \centering
\begin{tikzpicture}[scale=0.8, node distance=1cm, thick]
  \wotsgraph[above][1][0][2][3][1][2]
\end{tikzpicture}
\caption{
  A \WOTSP\ signature for message $m=(1,0,2,3)$ with checksum $c=(1,2)$.
  $\hash$ and $\fash$ are hash functions, nodes represent hash function outputs. 
  The nodes that are composed to produce the signature are shown in \colorbox{brightgreen}{\color{white!85!brightgreen}green}.
  The public key is a hash of the chain endpoints.
  Here, chain length $w = 4$, message length $\ell_1 = 4$, and checksum length $\ell_2 = 2$.
}\label{fig:wotsp_sig}
\end{figure}
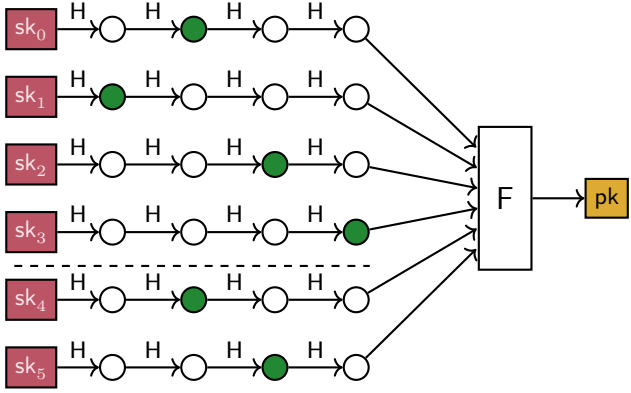

\subsubsection{The Winternitz One-Time Signature Scheme\texorpdfstring{\textsuperscript{+}}{+}}
The \acf{WOTSP} is a hash-based one-time signature scheme.
It consists of a series of hash chains, for which the starting points, together, can be considered to be equivalent to the \WOTSP{} secret key.
To sign a message, a hash function is applied to each starting point a message-dependent number of times, as depicted in \Cref{fig:wotsp_sig}.
The message to be signed is appended with a checksum on this message, to avoid, e.g., an attacker to forge a signature on $m_1 = (1, 1, 2, 3)$ given the signature on $m_0 = (1, 0, 2, 3)$.
A \WOTSP\ instance is characterized by the security parameter $\secpar$, indicating the output size of the hash function, and the message block size $\lgw$, which determines the chain length $w=2^{\lgw}$.
Derived from $\lgw$ and $n$ are the number of message chains $\ell_1$ and checksum chains $\ell_2$. 

\begin{figure}
    \centering
    \begin{tikzpicture}[level distance=1.5cm,
        level 1/.style={sibling distance=4cm},
        level 2/.style={sibling distance=2cm},
        auth/.style={rectangle,draw=brightblue,thick}]
        \node {$\pk^X = \hash(\sfh_{1,2} | \sfh_{3,4})$}
                child {node {$\sfh_{1,2} = \hash(\sfh_1 | \sfh_2)$}
                    child {node {$\sfh_1 = \hash(\pk_1^W)$}}
                    child {node[auth] {$\sfh_2$}}
                }
                child {node[auth] {$\sfh_{3,4}$}
                    child {node {$\sfh_3$}}
                    child {node {$\sfh_4$}}
                };
    \end{tikzpicture}
    \caption{
        A Merkle tree of height 2 with $2^2 = 4$ leaves.
        The leaf $\sfh_i$ represents the hash of a \WOTSP\ key $\pk_i^W$.
        A parent node $\sfh_{i,j}$ represents the hash of the concatenation of its children $\sfh_i, \sfh_j$.
        The final hash is the root node $\pk^X$ and represents the public key.
        The nodes forming the authentication path $\auth{\pk_1^W}$ are highlighted with \fcolorbox{brightblue}{white!0}{blue} rectangles.
        }\label{fig:xmss_tree}
\end{figure}
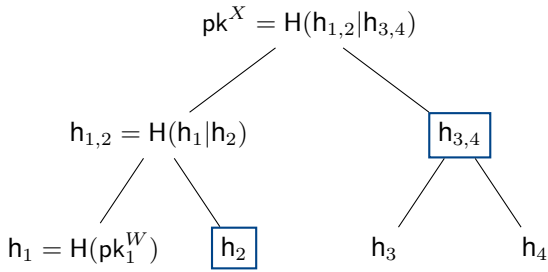

\acreset{XMSS}
\subsubsection{\ac{XMSS}} 
\XMSS\ is a hash-based signature scheme that is able to sign more than one message by organizing a large number of \WOTSP{} key pairs as a Merkle tree (\cref{fig:xmss_tree}).
The \WOTSP{} public keys are the leaves of the Merkle tree.
An \XMSS{} signature on a message consists of the \WOTSP{} signature on the message, and the authentication path that allows a verifier to recompute the Merkle tree root from the \WOTSP{} public key that was used, confirming its legitimacy.
Although \XMSS{} can sign more than one message, it must be ensured that the same \WOTSP{} instance is not used to sign more than one message, i.e., \XMSS{} is a \emph{stateful} scheme.
An \XMSS\ instance is characterized by the security parameter $\secpar$ and tree height $h'$, implying $2^{h'}$ \WOTSP\ key pairs as leaves.

\XMSS{} instances can be organized in a hierarchical fashion as \emph{hypertrees}, or tree of trees.
In this case, the Merkle tree root (and hence the public key) of \XMSS{} level $l$ is signed by one of the leaves of \XMSS{} level $l+1$.

\begin{figure}[t]
  \begin{subfigure}[b]{\linewidth}
    \centering
    \begin{tikzpicture}
      \node (dots) {$\dots$};
      \wotssig[$\sig_{l^*}^W$]{i}{dots}
      \xmss
        [\color{\graftingred}$\hat{\pk}_{l^*}^X$]
        [$\auth{\pk_{l^*}^W}$]
        {i}
        {wi}
      \node at ([xshift=4mm,yshift=4mm]xmsstri) {\color{\graftingred} \huge \Lightning};
      \wotssig[{\color{\graftingred}$\hat{\sig}_{l^*+1}^W$}]{i+1}{xmssi}
      \xmss
        [$\pk_{l^*+1}^X$]
        [$\auth{\pk_{l^*+1}^W}$]
        {i+1}
        {wi+1}
      \draw[->] (dots) -- (wi);
      \draw[->] (xmssi) -- (wi+1);
    \end{tikzpicture}
    \caption{Faulting \XMSS\ computation.}
    \label{fig:grafting_tree_attack_fault}
  \end{subfigure}%
  
  \begin{subfigure}[b]{\linewidth}
    \centering
    \begin{tikzpicture}
      \begin{scope}[draw=\graftingred, fill=\graftingred, every node/.style={text=\graftingred}]
        \node (dots) {$\dots$};
        \wotssig[$\tilde{\sig}_{l^*}^W$]{i}{dots}
        \xmss
          [\color{\graftingred}$\tilde{\pk}_{l^*}^X$]
          [\color{\graftingred}$\auth{\tilde{\pk}_{l^*}^W}$]
          {i}
          {wi}
        \draw[->,draw=\graftingred] (dots) -- (wi);
        \draw[->] (dots) -- (wi);
      \end{scope}
      \wotssig[{\color{\graftingred}$\tilde{\sig}_{l^*+1}^W$}]{i+1}{xmssi}
      \draw[->,draw=\graftingred] (xmssi) -- (wi+1); 
      \xmss
        [$\pk_{l^*+1}^X$]
        [$\auth{\pk_{l^*+1}^W}$]
        {i+1}
        {wi+1}
    \end{tikzpicture}
    \caption{Grafting an \XMSS\ tree.}
    \label{fig:grafting_tree_attack_graft}
  \end{subfigure}%
  \caption{
    Grafting tree attack on \SLHDSA~\cite{genetProtectingSPHINCSFault2023}.
  }
  \label{fig:grafting_tree_attack}
\end{figure}
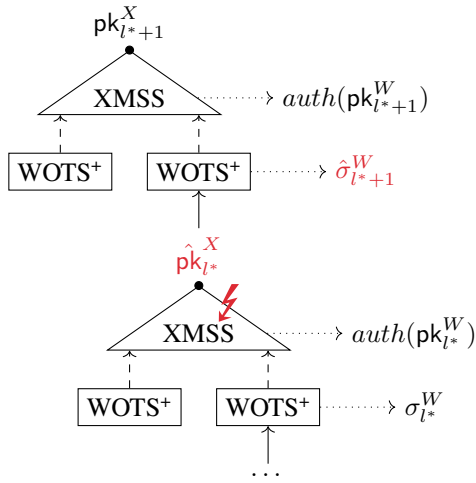
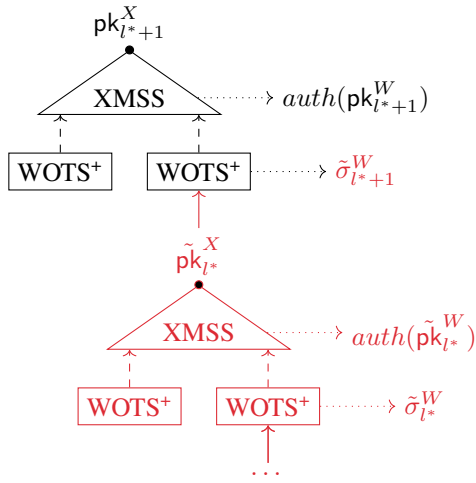

\subsubsection{SLH-DSA} 
\SLHDSA{} uses \FORS{} and \XMSS{} as building blocks to construct a practical scheme that can sign many different messages and is \emph{stateless}, i.e., \SLHDSA{} does not require keeping track of which secret keys were already used.

First, the message is signed using a \FORS{} key pair.
The \FORS{} signatures's public key is signed using an \XMSS{} hypertree.
The root node of the top-level hypertree corresponds to the public key of the scheme.
The particular \FORS{} key pair to use and the resulting path through the hypertree of \XMSS{} trees is determined by a value $R$, which is the output of a hash of the secret key and the message to be signed (deterministic mode), or the secret key, the message, and additional user-provided entropy (randomized mode).
An \SLHDSA{} signature consists of a \FORS\ signature of the message and a series of \XMSS{} signatures proving the \FORS{} key's validity.

\SLHDSA\ is standardized with the hash functions SHA\nobreakdash-2 and SHAKE.
The standard describes both \emph{small} and \emph{fast} variants (trading off the number of \XMSS{} levels and tree size) and in different security categories (with, respectively, 128-bit, 192-bit, and 256-bit hash output lengths).
For the detailed specification of each variant, we refer to FIPS~205~\cite{nistStatelessHashBasedDigital2024}.

\subsection{Grafting Tree Attack}

Under correct operation, the design of \SLHDSA\ ensures that each \WOTSP\ secret key is only used to sign \emph{one single} public key root. 
This invariant is crucial to uphold, since the security of \WOTSP\ degrades rapidly as the number of signatures with the same secret key increases.
The \emph{grafting tree attack}~\cite{castelnoviGraftingTreesFault2018} builds on this observation.
By triggering faulty computations during signature generation, it causes the same \WOTSP\ secret key of a given \XMSS{} level to sign multiple different public key roots of previous \XMSS{} levels.
Hence, it exposes additional \WOTSP\ secret values (cf. \cref{fig:wotsp_sig}), rendering the \WOTSP{} instance \emph{compromised}.
In particular, during the computation of an \XMSS\ tree, the attacker injects faults such that the root of the tree gets corrupted (\cref{fig:grafting_tree_attack_fault}).
Note that any fault during the entire \XMSS{} computation leads to the corruption of its root.
The faulty root is then signed by the next level in the hypertree using \WOTSP.
When the process is repeated, the attacker can gather multiple signatures with the same \WOTSP\ secret key. 

The attacker can use a compromised \WOTSP{} instance as in \cref{fig:grafting_tree_attack_graft}.
With enough signatures, the attacker can use a compromised \WOTSP\ instance (in layer~$l^*+1$) to sign a \emph{grafted} (i.e., fully attacker-generated) \XMSS\ tree (in layer~$l^*$).
There is some offline computation effort to arrive at a grafted tree for which the public key is compatible with the values that are leaked from the compromised \WOTSP{} instance.
As each additional signature potentially reveals new \WOTSP\ chain values, obtaining more faulty signatures that visit the same \WOTSP{} key pair reduces the tree grafting complexity.
The most straightforward version of the attack uses a valid (uncorrupted) signature in addition to the faulted signature(s), but prior work showed that this is not a strict requirement~\cite{genetProtectingSPHINCSFault2023}.

From the point of view of the signature verifier, the grafted tree appears valid, as it contains a valid signature from the \WOTSP{} key pair in layer~$l^*+1$.
Since the grafted tree is fully attacker-controlled, including its own secret keys, an attacker uses a grafted tree to produce a forgery for any chosen message that visits the compromised \WOTSP{} node.

\section{Threat Model}
\label{sec:threat_model}

We adopt the standard Rowhammer attack model that differentiates between an offline phase and an online phase.
During the offline phase, the attacker has unfettered access to a replicated target system to prepare the attack.

In the online phase, the attacker's access is heavily restricted, as detailed in the following.
The attacker can execute code (e.g., \tech{Swage}) with user-level privileges on the target system.
They have access to high-resolution timers like \instr{rdtscp}. %

The target system uses DDR4 memory, possibly with \TRR{} mitigation.
The attacker is able to co-locate their code with the victim on the same logical CPU core~\cite{adilettaMayhemTargetedCorruption2024,kwongRAMBleedReadingBits2020,DBLP:conf/sp/RakinCYF22}.
Finally, we assume that the attacker can influence when the victim process allocates the memory for the targeted data structure, as, e.g., enabled by the fork-on-connection strategy used in security-focused server processes like OpenSSH~\cite{kwongRAMBleedReadingBits2020} or swapping due to memory pressure~\cite{DBLP:conf/sp/RakinCYF22}.

\section{Swage: A Modular Universal End-to-End Rowhammer Attack Framework}
\label{cha:swage}

\newcommand{\offlineicon}{\ding{109}}
\newcommand{\onlineicon}{\ding{108}}
\newcommand{\bothicon}{\ding{119}}

\begin{figure}
    \centering
    \resizebox{\linewidth}{!}{
    \begin{tikzpicture}[
    box/.style={rectangle, minimum width=1cm, minimum height=1cm, align=center, fill=brightblue!10, draw=brightblue!80},
    strategy/.style={minimum width=1cm, minimum height=1cm, align=center},
    node distance=2em
    ]

    \node[box] (dram) {DRAM Inspector\\\footnotesize(\cref{sec:dram_inspector})};
    \node[strategy, above right=1.5em and 5em of dram.center, anchor=west] (nodrama) {\offlineicon{} \tech{Drama}~\cite{pesslDRAMAExploitingDRAM2016}};
    \node[strategy, right=5em of dram.center, anchor=west] (xiao) {\offlineicon{} Xiao et al.~\cite{xiaoOneBitFlips2016}};
    \node[strategy, below right=1.5em and 5em of dram.center, anchor=west] (dram-config) {\onlineicon{} \tech{ConfigFile}};
    \node[box, below=of dram] (alloc) {Allocator\\\footnotesize(\cref{sec:swage_allocator})};
    \node[strategy, above right=0.75em and 5em of alloc.center, anchor=west] (hugepage) {\offlineicon{} \tech{HugePage}};
    \node[strategy, below right=0.75em and 5em of alloc.center, anchor=west] (spoiler) {\onlineicon{} \tech{Spoiler}~\cite{islamSPOILERSpeculativeLoad2019}};
    \node[box, below=of alloc] (hammerer) {Hammerer\\\footnotesize(\cref{sec:hammerer})};
    \node[strategy, above right=1.5em and 5em of hammerer.center, anchor=west] (blacksmith) {\offlineicon{} \tech{BlacksmithFuzz}};
    \node[strategy, right=5em of hammerer.center, anchor=west] (replay) {\onlineicon{} \tech{ConfigFile}};
    \node[strategy, below right=1.5em and 5em of hammerer.center, anchor=west] (devmem) {\offlineicon{} \tech{DevMem}};
    \node[box, below=of hammerer] (orchestrator) {Orchestrator\\\footnotesize(\cref{sec:orchestrator})};
    \node[box, below=of orchestrator] (injector) {Page Injector\\\footnotesize(\cref{sec:page_injection})};
    \node[strategy, right=5em of injector.center, anchor=west] (mayhem) {\bothicon{} \tech{FrameFengShui}~\cite{kwongRAMBleedReadingBits2020}};
    \node[box, below left=2em and 4em of orchestrator] (process) {Victim}; %

    \draw[->] (dram) -- node[midway,left] {\small Address Mapping} (alloc);
    \draw[->] (alloc) -- node[midway,left] {\small Memory} (hammerer);
    \draw[->] (hammerer) -- node[midway,left] {\small Hammering Primitive} (orchestrator);
    \draw[->] (orchestrator) -- node[midway,right] {\small Launches} (injector);
    \draw[->] (injector) -- node[midway,below] {\small Injects} (process);
    \draw[<->] (process) -- node[midway,above left] {\small Communicates} (orchestrator);

    \draw[->, loop left, looseness=2] (hammerer) to node[midway, left]{\small Profiling} (hammerer);

    \draw[->] (nodrama.west) -- (dram);
    \draw[->] (xiao.west) -- (dram);
    \draw[->] (dram-config.west) -- (dram);
    \draw[->] (hugepage.west) -- (alloc);
    \draw[->] (spoiler.west) -- (alloc);
    \draw[->] (blacksmith.west) -- (hammerer);
    \draw[->] (replay.west) -- (hammerer);
    \draw[->] (devmem.west) -- (hammerer);
    \draw[->] (mayhem.west) -- (injector);
    \end{tikzpicture}}
  \caption{Flowchart showing the steps of a Rowhammer attack using \tech{Swage}. \offlineicon{} depicts offline phase usage of the strategy, \onlineicon{} stands for online usage and \bothicon{} is a strategy used in both offline and online phase of an attack.}
  \label{fig:swage_flowchart}
\end{figure}
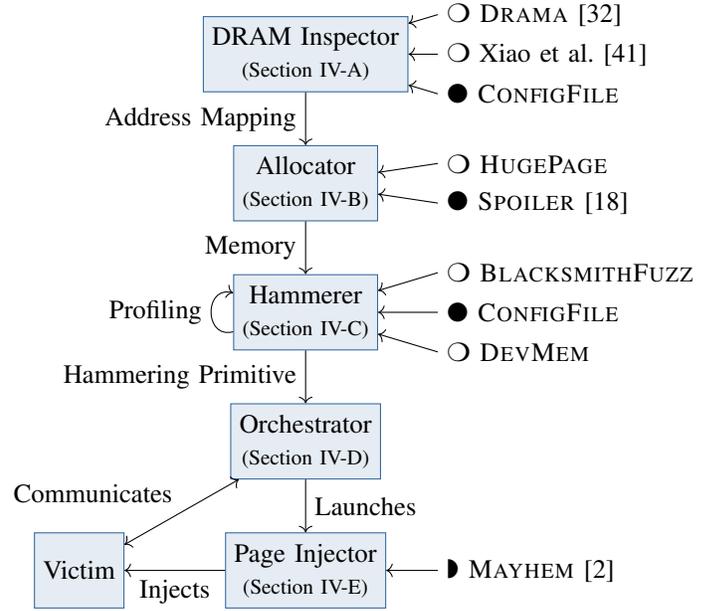

Conducting Rowhammer attacks in practice is challenging:
As the attack interacts with different aspects of the target system, many moving parts need to be considered.
Though a number of tools have been proposed to facilitate Rowhammer attacks, they often cover only a specific aspect of the attack, such as finding reproducible hammering patterns or reverse-engineering the physical memory layout~\cite{frigoTRRespassExploitingMany2020,pesslDRAMAExploitingDRAM2016,jattkeBlacksmithScalableRowhammering2022,bolcskeiRubiconPreciseMicroarchitectural2025}.
Most of the time, the software design of these tools is tightly coupled to the evaluation scenario of the respective paper, making it difficult to reuse them in an end-to-end Rowhammer attack.
We propose \tech{Swage}, a novel modular end-to-end framework for Rowhammer attacks.
\tech{Swage} is implemented in Rust, C, and Python, and is designed to be modular and extensible.
We provide a \tech{Swage} artifact as an open-source project and invite the community to contribute to its development.

\cref{fig:swage_flowchart} shows the modular architecture of \tech{Swage}.
Most modules have multiple instantiation options that allow choosing between easy prototyping of attacks by using root privileges or performing realistic attacks that only require user space access.
In the following, we describe each module and its role in the attack workflow in more detail.

\subsection{DRAM Inspector}\label{sec:dram_inspector}
In order to conduct a Rowhammer attack, the attacker needs information about the locality in memory to arrange the aggressor rows and the victim row for their attack.
The \tech{Dram Inspector} module provides information about the physical memory layout of the target system.
When accessing a memory location, the memory controller applies a \emph{memory mapping} function to the physical address.
This function is specific to any given CPU generation and usually not disclosed by manufacturers.
However, there exist side-channel attacks that can be used to reverse-engineer the physical memory layout.
For this, \tech{Swage} employs a software suite combining the outputs of both \tech{Drama}~\cite{pesslDRAMAExploitingDRAM2016} and Xiao et al.~\cite{xiaoOneBitFlips2016} for improved reliability.
This step only needs to be done once per targeted CPU during the offline phase.
Afterward, the mapping can be loaded from a configuration file.

\subsection{Allocator}\label{sec:swage_allocator}
To construct the memory access patterns required for Rowhammer, the attacker needs to be able to allocate physically contiguous memory.
The \tech{Allocator} module provides a unified interface for allocating contiguous memory blocks.
\tech{Swage} comes with two implementations for allocating these blocks, supporting both attack prototyping with elevated privileges and realistic scenarios.
For the former, the attacker can configure the Linux kernel to reserve memory for huge pages at boot time and then allocate them via the \tech{HugePage} instantiation of the \tech{Allocator} module to get access to vast amounts of physically contiguous memory as demonstrated by prior work~\cite{jattkeBlacksmithScalableRowhammering2022}.

For realistic scenarios, the attacker has access to the \tech{Spoiler} instantiation.
While the Linux allocator does not make any guarantees on physical contiguous memory for larger chunks of allocated memory, its use of the buddy allocator strategy makes it highly likely that some parts of the allocated memory are indeed physically contiguous~\cite{kwongRAMBleedReadingBits2020}.
Following~\cite{adilettaMayhemTargetedCorruption2024}, we use the Spoiler side-channel~\cite{islamSPOILERSpeculativeLoad2019} to reveal all 1 MiB blocks of contiguous memory inside our allocated buffer.
Note that their approach does not give information on the positioning of these 1 MiB blocks relative to each other.
As discussed later, 1 MiB of contiguous memory may not always suffice to find a hammering pattern.
Thus, we use the row buffer side-channel in combination with our already obtained information about the \DRAM{} addressing function to reveal which of the discovered 1 MiB blocks are contiguous and form 4 MiB blocks of contiguous memory.

\subsection{Hammerer}\label{sec:hammerer} 
The \tech{Hammerer} module is the core of \tech{Swage} and implements strategies to fuzz and replay Rowhammer patterns.
The \tech{BlacksmithFuzz} instantiation is a fork of the Blacksmith fuzzer~\cite{jattkeBlacksmithScalableRowhammering2022}.
It is intended to be used in conjunction with the \tech{HugePage} memory allocation strategy to find reproducible hammering patterns during the offline phase of the attack.
Our analysis shows that Blacksmith tends to generate patterns that require large amounts of physically contiguous memory, making them unsuitable in a realistic attack scenario.
Thus, we added a novel component that breaks up the hammering pattern into chunks that only require 1 or 4 MiB of contiguous memory and re-evaluates their reproducibility. 
Rapid attack prototyping in the offline phase can be achieved with the \tech{DevMem} backend:
Instead of performing Rowhammer, it allows to architecturally flip a bit in the targeted location by using the Linux kernel's \texttt{/dev/mem} interface to access arbitrary physical memory.
In the online phase, the \tech{ConfigFile} backend is used to load and hammer a previously found pattern.

\subsection{Orchestrator}\label{sec:orchestrator}
The \tech{Orchestrator} module implements the actual attack against a victim process.
As input from the previous stages, it has access to a hammering primitive and a mapping to the memory page $\mathcal{P}$ at which the hammering primitive will cause a bit flip.
The details of the victim process are abstracted by a victim-specific wrapper that needs to implement four methods: \texttt{start}, \texttt{init}, \texttt{check}, \texttt{stop}.
To start the attack workflow, the \tech{Orchestrator} in parallel calls the \texttt{start} method and the page injector.
The \texttt{start} method may be used to trigger the creation of the victim process or other initialization steps.
The page injector ensures that the victim uses the vulnerable page $\mathcal{P}$ to store the data that the attacker wants to manipulate via Rowhammer and will be discussed in \cref{sec:page_injection}. %
Afterward, the \tech{Orchestrator} calls the \texttt{init} method to synchronize with the execution of the victim and start the Rowhammer attack. 
The synchronization step could, e.g., involve triggering the victim via a network request or the use of a cache attack.
Next, the \texttt{check} method is called to ascertain the attack's success, assuming the victim creates some kind of output or otherwise observable behavior.
Finally, the \texttt{stop} method terminates the attack cycle.

\subsection{Page Injection}\label{sec:page_injection}
\emph{Page injection} refers to techniques to manipulate the Linux memory allocator to return a specific, attacker-chosen memory page upon the next allocation.
In the context of Rowhammer, this enables the attacker to ensure that the victim places the data structure \emph{exactly} on the page that the attacker can introduce bit flips in.
\tech{Swage} implements a technique from prior work~\cite{kwongRAMBleedReadingBits2020,adilettaMayhemTargetedCorruption2024} that exploits an optimization in the Linux kernel allocator.
The optimization consists of a per-CPU list of memory pages that is consulted during memory allocation before going to the main memory allocator.
This can be exploited to inject a target page into a victim process co-located on the attacker's core.
A recent novel page injection technique, \tech{Rubicon}~\cite{bolcskeiRubiconPreciseMicroarchitectural2025}, combines page injection and eviction strategies to implement cross-core page injection.

\section{Evaluation of \tech{Swage}}

During the offline part of a Rowhammer attack, the attacker collects information about the target.
We evaluated the components of \tech{Swage}, experimentally verifying the reliability of the attack primitives presented in \cref{cha:swage}.
For evaluating an attack on \SLHDSA{},  we conducted our \tech{Swage} experiments on a system with an Intel i5-6400 CPU and a single G.Skill AEGIS 16 GB DDR4-2133 DIMM, running Ubuntu 20.04.6 with Linux kernel version 6.8.12-generic with default settings.
We target an implementation with attackable data residing on the stack; the concrete instantiations of such data are discussed in \cref{sec:slashdsa}.
In the following, we describe the results of our \tech{Swage} analysis.

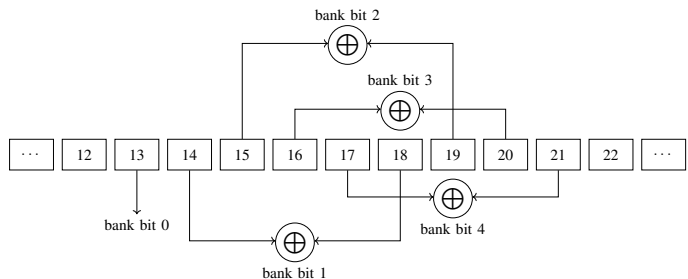
\begin{figure}
    \centering
    \resizebox{\linewidth}{!}{
    \begin{tikzpicture}[
        bit/.style={draw,rectangle,minimum width=1cm,minimum height=0.7cm},
        xor/.style={draw,circle,inner sep=1.5pt,font=\Large} 
      ]
    
      \node[bit] (b11) at (11*1.2,0) {$\dots$};
      \foreach \i in {12,...,22}{
        \node[bit] (b\i) at (\i*1.2,0) {\i};
      }
      \node[bit] (b11) at (23*1.2,0) {$\dots$};
    
      \draw[->] (b13.south) -- ++(0,-1) node[below] {bank bit 0};

      \foreach \a/\b/\bankIdx/\yxor/\xorlblpos in {
        14/18/1/-2/below,
        15/19/2/2.5/above,
        16/20/3/1/above,
        17/21/4/-1/below
      }{
        \pgfmathsetmacro{\xm}{(\a+\b)/2*1.2}
        \node[xor,label=\xorlblpos:{bank bit \bankIdx}] (x\bankIdx) at (\xm,\yxor) {$\bigoplus$};
        \draw[->] (b\a) |- ([yshift=1pt]x\bankIdx.west);
        \draw[->] (b\b) |- ([yshift=1pt]x\bankIdx.east);
      }
    \end{tikzpicture}}
\caption{
    \DRAM\ bank-selection mapping for the target system.
    Bank bit 0 is taken directly from physical bit 13, while bank bits 1-4 are computed as the XOR ($\oplus$) of bit pairs ($b_{14} \oplus b_{18}$, $b_{15} \oplus b_{19}$, $b_{16} \oplus b_{20}$, and $b_{17} \oplus b_{21}$), respectively. 
    Dots indicate omitted neighboring physical address bits.
}\label{fig:bank_addrs}
\end{figure}

\subsection{Reverse-Engineering the \DRAM\ Address Mapping}

For a Rowhammer attack, an attacker needs to determine the mapping of physical addresses to \DRAM{} banks, rows, and columns.
\tech{Swage} uses the row conflict timing measurements to determine the physical mapping function 
running the algorithms from both \tech{Drama}~\cite{pesslDRAMAExploitingDRAM2016} and Xiao et al.~\cite{xiaoOneBitFlips2016} for improved reliability.
The row mapping function for the target system is simple: the row index is determined by physical bits 18 to 29.
Similarly, the column index is determined by physical bits 0 to 12.
\cref{fig:bank_addrs} shows the reverse-engineered bank mapping function for the target system.
We verify the findings by checking the timing of pairs of addresses with same or different bank indices.
As it shows no significant outlier behavior, the mapping function is considered correct.

\subsection{Contiguous Memory Allocation}

We evaluated the \tech{Spoiler} implementation in \tech{Swage} by repeatedly allocating a \MiB{4} block using the \tech{Allocator} module.
We configure the allocator to search for contiguous memory blocks in a \GiB{2} buffer.
Repeating the allocation experiment 100 times, we measured the elapsed time for the memory allocation and the false-positive rate, i.e., the number of times the allocator falsely identifies a physically contiguous \MiB{4} block via the \tech{Spoiler} timing side-channel.
On the target machine, we find that an allocation using \tech{Spoiler} takes 41.6 seconds on average, with a false positive rate of 66\%.
However, the falsely identified \MiB{4} memory blocks consist of either a pair of \MiB{2} blocks, a quartet of \MiB{1} blocks, or one \MiB{3} block and one \MiB{1} block.
While technically non-contiguous, those \emph{composite blocks} can still be used in a Rowhammer attack if they are consistent with the bank timing function, i.e., if they show the expected timing measurements due to row buffer conflicts.

\subsection{Page Injection}
Evaluating the page injection technique in the \tech{Swage} \tech{Page Injector}, we find that the page injection is successful in most of the cases.
This is in line with findings in previous work~\cite{kwongRAMBleedReadingBits2020,amerPQHammerEndtoEndKey2025}.
When the page injection fails, the attacker restarts the online phase of the attack and tries again.
Those failed page injection attempts most likely happen due to noise caused by allocations from different processes on the same CPU. 
As soon as the page injection is successful, the attacker can proceed to collect signatures from the signing server.

\subsection{Reproducible Hammering Patterns}

\tech{Swage} employs \tech{Blacksmith}~\cite{jattkeBlacksmithScalableRowhammering2022} to find reproducible hammering patterns in the \tech{Hammerer} module.
\tech{Blacksmith} is a fuzzing-based software suite to find reproducible hammering patterns.
It generates \emph{non-uniform access patterns} by randomizing three domains:
How often an aggressor row is activated (frequency), 
the time between the start of the hammering pattern and the first activation of an aggressor row (phase), 
and how often an aggressor row is activated back-to-back (amplitude).
By using \tech{Blacksmith}, \tech{Swage} can effectively bypass \TRR, enabling Rowhammer attacks in state of the art DDR4 memory modules.
The hammering pattern from \tech{Blacksmith} can be imported in \tech{Swage}, where the pattern is just-in-time compiled to an assembly function to maximize the hammering performance.

\begin{table}
  \centering
  \caption{
    Reproducibility results of \tech{BlacksmithFuzz}.
    Each candidate pattern was tested over 10 rounds, each allowing up to 1000 pattern repetitions or terminating early upon the first encountered bit flip.
  }\label{tab:blacksmith_repro_results}
  \resizebox{\linewidth}{!}{\begin{tabular}{l@{\hspace{2em}}rr@{\hspace{2em}}rr@{\hspace{2em}}rr}
\toprule
Pattern & \multicolumn{2}{c@{\hspace{2em}}}{Bit Flips} & \multicolumn{2}{c@{\hspace{2em}}}{Retries} & \multicolumn{2}{c}{Time (s)} \\
\cmidrule(l{0.125em}r{2em}){2-3}\cmidrule(lr{2em}){4-5}\cmidrule(r){6-7}
 & Avg. & Max & Avg. & Max & Avg. & Max \\
\midrule
$\mathcal{A}$ & 1.3 & 2 & 1.4 & 4 & 1.22 & 3.54 \\
$\mathcal{B}$ & 0.0 & 0 & 1000.0 & 1000 & 901.10 & 901.45 \\
$\mathcal{C}$ & 0.0 & 0 & 1000.0 & 1000 & 883.49 & 885.65 \\
$\mathcal{D}$ & 2.1 & 6 & 8.9 & 26 & 7.93 & 23.13 \\
$\mathcal{E}$ & 1.1 & 2 & 271.6 & 780 & 236.13 & 678.13 \\
$\mathcal{F}$ & 0.2 & 1 & 863.9 & 1000 & 774.79 & 896.88 \\
$\mathcal{G}$ & 1.0 & 1 & 23.7 & 51 & 20.23 & 43.53 \\
$\mathcal{H}$ & 1.1 & 2 & 10.8 & 34 & 9.52 & 29.97 \\
\bottomrule
\end{tabular}
}
\end{table}

We evaluated \tech{BlacksmithFuzz} on the target machine, summarizing the results of an eight hour fuzzing run in \cref{tab:blacksmith_repro_results}.
The table shows reproducibility results for eight different hammering patterns showing at least one bit flip in the fuzzing run.
We see from the table a varying degree of reproducibility:
While patterns $\mathcal{B}$ and $\mathcal{C}$ are not reproducible at all, patterns $\mathcal{A}$, $\mathcal{D}$, and $\mathcal{H}$ are highly reproducible.
However, pattern $\mathcal{D}$ usually causes bit flips in multiple locations in \DRAM.
While this can be a desired property for some attack scenarios, the rogue flips potentially lead to system instability.
Therefore, for the SLasH-DSA attack, we chose pattern $\mathcal{G}$ due to its high reproducibility and controlled bit flipping behavior.

\subsection{Reproducibility of Hammering Patterns With Block Splitting}
As discussed in \cref{sec:hammerer}, \tech{Swage} needs to adapt a hammering pattern found with \tech{BlacksmithFuzz} for use with distributed memory blocks.
Block splitting allows the attacker to split the aggressor mapping of an access pattern into smaller memory blocks eliminating the need for vast amounts of physically contiguous memory.
For the splitting process, the 1 GiB huge page of physically contiguous memory used by \tech{BlacksmithFuzz} is divided into \MiB{4} chunks and corresponding \MiB{4} memory regions are allocated -- using \tech{Spoiler} in this case -- on the target system.
This process preserves the local properties of the patterns withing each \MiB{4} chunks but destroys the global pattern, as the individual \MiB{4} blocks are (w.h.p.) not adjacent to each other in physical memory.
However, as gathering a large amount of \MiB{4} blocks using the \tech{Spoiler} approach is time-consuming, we also limit the size of the patterns that the fuzzer is allowed to generate.
For the pattern $\mathcal{G}$ that we selected in the preceding subsection, we require a total of ten \MiB{4} chunks, each containing a subset of the aggressors. 
The reproducibility experiment confirms that the pattern remains effective even when applied to these smaller memory blocks.

\subsection{Combined \tech{Swage} Validation}
In a last offline reproducibility experiment, the attacker deallocates the target page before initiating the hammering process, and then allocates it in a \emph{dummy process}.
This dummy process reserves a large stack-based array, fills it with a predetermined value, and then reads the array to verify whether the contents of the target page match the expected value before rewriting the array.
This approach mimics the online phase of the attack, where a victim process is anticipated to interact with the target page.
The results of this assessment demonstrate that the pattern maintains the reproducibility outlined in \cref{tab:blacksmith_repro_results}, underscoring its resilience to page walks, page table flushes, and other memory management operations carried out by the operating system.
With that, \tech{Swage} has successfully determined a reproducible access pattern and aggressor mapping that can then be used to induce bit flips in the target page.

\section{SLasH-DSA Attack}\label{sec:slashdsa}

We now present the SLasH-DSA attack, a novel end-to-end Rowhammer attack against SLH-DSA.
We take previous work by Gen\^et~\cite{genetProtectingSPHINCSFault2023} as a starting point for our discussion and refer to their work for further insights into complexity considerations not covered in this work.

Our work extends Gen\^et's attack in several key aspects:
While Gen\^et demonstrated the tree grafting attack using hardware-based clock glitching on embedded systems, we show for the first time that the attack is feasible with software-only Rowhammer on commodity hardware.
In his practical experiments, Gen\^et targeted the SHAKE-256s parameter set in a modified \SPHINCSP{} reference implementation, blindly injecting clock glitches while computing \XMSS{} trees and \WOTSP{} signatures.
In contrast, we target all parameter sets in OpenSSL's implementation of the now-standardized \SLHDSA{}, specifically attacking the \var{lnode} buffer in the \func{xmss\_node} function during \XMSS{} tree computation (\cref{sec:openssl_fault_analysis}).
This allows us to exploit the extended lifetime of \var{lnode} on the stack, providing larger temporal windows for Rowhammer-induced bit flips.
Beyond the attack vector and target, we introduce a novel exact complexity analysis for concrete compromised \WOTSP{} instances, enabling optimal candidate selection, whereas Gen\^et analyzed the average case complexity for uniformly distributed faults.

We first describe the fault analysis of the OpenSSL \SLHDSA{} implementation, then continue with the signature generation phase, where we leverage the \tech{Swage} framework to carry out the attack, and finally describe the post-processing phase with the tree grafting attack.

\begin{algorithm}
\caption{$\text{xmss\_node}(\sk, i, z, \pk, ADRS)$ \cite{nistStatelessHashBasedDigital2024}}
\label{alg:xmss_node}
\DontPrintSemicolon
\KwIn{Secret seed $\sk = \SKseed$, target node index $i$, target node height $z$, public seed $\pk$, address $ADRS$}
\KwOut{$n$-byte root $node$}

\If{$z = 0$}{
    $ADRS.\text{setTypeAndClear}(\text{WOTS\_HASH})$\;
    $ADRS.\text{setKeyPairAddress}(i)$\;
    $node \leftarrow \text{wots\_pkGen}(\sk, \pk, ADRS)$\;
}
\Else{
    $lnode \leftarrow \text{xmss\_node}(\sk, 2i, z - 1, \pk, ADRS)$\;
    $rnode \leftarrow \text{xmss\_node}(\sk, 2i + 1, z - 1, \pk, ADRS)$\;
    $ADRS.\text{setTypeAndClear}(\text{TREE})$\;
    $ADRS.\text{setTreeHeight}(z)$\;
    $ADRS.\text{setTreeIndex}(i)$\;
    $node \leftarrow H(\pk, ADRS, l \| r)$\;
}
\Return $node$\;
\end{algorithm}

\newcommand{\xmssNode}{\func{xmss\_node}}

\subsection{Fault Analysis of the OpenSSL \SLHDSA\ Implementation}\label{sec:openssl_fault_analysis}
In the hypertree phase of the \SLHDSA\ signature generation, \xmssNode\ (\cref{alg:xmss_node}) is called repeatedly to compute the root nodes of the Merkle trees identified by addresses \ADRS, a public seed $\pk$, and a secret $\sk$.
In OpenSSL, \xmssNode\ is implemented in file \file{crypto/slh\_dsa/slh\_xmss.c}.
It takes as input the secret and public seed, the start index, the target height, and the \ADRS\ of the current tree.
The function computes the root node of the Merkle tree bottom up, starting at the left-most \WOTSP\ node and working its way up to the root, closely matching the recursive depth-first search algorithm from the standard.
The function recursively calls itself with the subtree corresponding to the current node's children until the bottom of the tree is reached.
Once at the bottom, the two leftmost \WOTSP\ public keys are computed from the secret seed, and the resulting public keys are stored in \var{lnode} and \var{rnode}.
After returning control flow to the parent, the siblings in \var{lnode} and \var{rnode} are concatenated, hashed, and passed to the grandparent.
This procedure is repeated until the top of the tree, where the root node acts as the scheme's public key.

\begin{figure*}
\begin{center}    
  \includegraphics[width=0.95\linewidth]{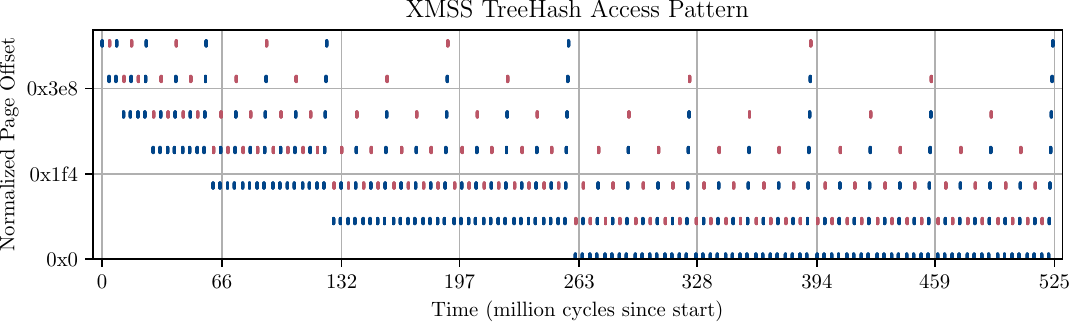}
    \caption{
        Access patterns for  \var{lnode} in the \func{ossl\_slh\_xmss\_node} function in the OpenSSL \SLHDSA\ implementation.
        The figure shows all accesses to the \var{lnode} buffers across all tree levels during generation of an \XMSS\ public key with the SHA2-256s parameter set.
        Write accesses are shown in \colorbox{brightred}{\color{white}red}, read accesses in \colorbox{brightblue}{\color{white}blue}.
    }\label{fig:slh_dsa_access_pattern_xmss}
\end{center}
\end{figure*}

We analyze the implementation of \xmssNode\ in OpenSSL to find suitable memory regions for our Rowhammer attack.
Specifically, we are interested in high spatial suitability,
i.e., the buffers in question are large enough to provide flexibility in the exact bit to be flipped, 
and high temporal suitability,
i.e., the memory location provides a large enough temporal window in which the bit flip is able to occur.
For spatial suitability, we observe that \var{lnode} holds the $n$-byte-output of hash function $\hash$.
When computing the topmost \var{lnode} in a tree, the algorithm descends into $h'-1$ levels, each maintaining its own \var{lnode}.
In total, this makes a memory region of $n \cdot (h'-1)$ bytes.
Note that the spatial suitability increases with higher security levels, as $n$ is 32 bytes for security category five, but only 16 bytes for security category one.

For the temporal suitability of \var{lnode},
we assume the signing party currently executes \xmssNode\ to compute the public key $\pk^X$ of an \XMSS\ tree and just compute \var{lnode} in line 6 at the topmost layer of the tree. 
From the pseudocode, we see that the left subtree of the current node is processed before the right one, similar to a depth-first search.
Meanwhile, the topmost \var{lnode} remains on the stack for a significant portion algorithm's runtime, which makes it a prime target for a Rowhammer attack.
We see that the temporal suitability decreases at lower levels in the \XMSS\ tree.
Therefore, the larger the size of the \XMSS{} tree, the lower the temporal precision of inducing the bit flip needs to be.
This makes the ``fast'' variants of SLH-DSA less temporally suitable.
In an attack, this issue can be tackled by applying performance degradation techniques discussed in \cref{sec:performance_degradation}.

\cref{fig:slh_dsa_access_pattern_xmss} shows the results of an experimental verification of this observation, showing the access patterns of \var{lnode} while computing an \XMSS\ tree in an instrumented build of OpenSSL.
After a warm-up phase for the recursive calls to saturate the layers of the tree, the topmost page offsets in the figure correspond to the accesses to \var{lnode} in the top layer, and lower page offsets are accesses in lower layers.
We observe that lower offsets in the array are written to and read from frequently, the higher offsets are accessed only few times per layer.
We conclude that \var{lnode} is a both spatially and temporally suitable target for a Rowhammer attack.

The information gathered in the fault analysis of OpenSSL's \SLHDSA\ implementation guides the implementation of the SLasH-DSA orchestrator module in \tech{Swage}.
This module incorporates attack-specific code and framework configuration for SLasH-DSA, while allowing to reuse \tech{Swage}'s core components described in \cref{cha:swage}.

\subsection{Performance Degradation \& Signature Collection}\label{sec:performance_degradation} 
After a fault analysis of OpenSSL, the attacker starts the online phase of the attack, where they apply memory massaging techniques to inject a target page into the victim process.
They then hammer the target page while continuously collecting signatures from the signing server until a timeout is reached.

\paragraph{Performance Degradation.}
If available, OpenSSL utilizes hardware acceleration and vector instructions when signing messages with \SLHDSA.
This increase in performance is most notable for the lower security levels of SHA-2 and \emph{fast} variants of \SLHDSA.
To still allow Rowhammer-induced bit flips against these variants, the signing server is slowed down using a \emph{performance degradation attack}, artificially slowing down the execution of a target program, a technique also employed by related work~\cite{amerPQHammerEndtoEndKey2025,adilettaLeapFrogRowhammerInstruction2025}.
In SLasH-DSA, the attacker occupies the victim's core by using the \var{stress} tool, which spawns a parameter-dependent number of CPU intensive worker threads.
Depending on the parameter set, they use as little as 15 workers for the ``small'' variants of \SLHDSA{} and 63 workers for the ``fast'' variants.
In scenarios where this amount of additional CPU load is unacceptable, performance degradation attacks like \tech{HyperDegrade}~\cite{aldayaHyperDegradeGHzMHz2022} or \tech{Controlled Preemption}~\cite{zhuControlledPreemptionAmplifying2025} pose viable alternatives.

\paragraph{Collecting Signatures.}
The attacker now collects signatures from the signing server while concurrently hammering the target memory region.
Due to the performance degradation attack, signing times are significantly prolonged.
In our evaluation, depending on the parameter set, signature collection rates range from approximately 95 to 1917 signatures per hour.
Given the spatial locality of Rowhammer-induced bit flips within DRAM rows, the attacker seeks corrupted signatures where faults affect the same \WOTSP{} instance across multiple signature attempts, as these enable the tree grafting attack described below.
The \tech{Swage} framework orchestrates this process, coordinating the Hammerer module's memory access patterns with signature collection until the configured timeout is reached.

After collecting signatures from the victim process, the attacker performs the grafting tree attack in an offline post-processing phase.
In that phase, the attacker searches for \XMSS\ trees matching the compromised \WOTSP\ instances (cf. \cref{background:slhdsa}) they collected from the victim (see \cref{fig:grafting_tree_attack}).
The methodology for post-processing described below was first introduced by Castelnovi et al. \cite{castelnoviGraftingTreesFault2018}.

\newcommand{\chain}{\func{chain}}

\subsection{Identifying Secret \WOTSP\ Values}\label{sec:identifying_secrets}
In this step, the attacker identifies \WOTSP\ secret values from the collected signatures.
First, the \WOTSP\ signatures (\cref{fig:wotsp_sig}) are extracted from the collected signatures and grouped by \ADRS.
The attacker then identifies the signatures for which the corresponding addresses map to the same \WOTSP\ instance.
Subsequently, they calculate the exposed \WOTSP\ secret values for each \ADRS.
Let $(\wsig, \hat{\sig}{}^W)$ be two distinct \WOTSP\ signatures for the same \WOTSP\ instance at \ADRS\ with target layer $l^*$, and let $m_i$ be the message chunk corresponding to the message signed by $\wsig$ in chain~$i$.
The attacker identifies \WOTSP\ secret values in $\hat{\sig}{}^W = (\hat{\sig}{}^{(0)}, \dots, \hat{\sig}{}^{(\ell-1)})$ using the exhaustive search described by Castelnovi et al.~\cite{castelnoviGraftingTreesFault2018}:
\begin{enumerate}
    \item For each $\hat{\sig}{}^{(i)} \in \hat{\sig}$: find $1 \leq k < m_i$ such that $$\chain\left(\hat{\sig}{}^{(i)}, k, w-1-k, \pk, \ADRS\right) = p_i.$$
    \item If such $k$ exists, then $\hat{\sig}{}^{(i)}$ exposes the \WOTSP\ secret values $k, \dots, m_i-1$ of chain $i$. If no $k$ leads to a match, the \WOTSP\ signature is discarded.
\end{enumerate}
\emph{Complexity}: Identifying secret \WOTSP\ values only yields a negligible computational cost, ranging from $2^{8.04}$ to $2^{12.98}$ hashes, depending on the parameter set.

\subsection{Tree Grafting}\label{sec:tree_grafting}
After identifying the \WOTSP\ secret values, the attacker performs the grafting tree attack.
In this step, the attacker tries to find an \XMSS\ public key that can be signed using the targeted \WOTSP\ instance.
This involves an exhaustive search for a secret $\hatSKseed$ that produces an \XMSS\ public key $\hat{\pk}{}^X$.
However, only a limited number of \XMSS\ trees can be signed, as the attacker can only sign messages that are compatible with the exposed chain secrets.
Let $(\hat{\theta}_0, \hat{\theta}_1, \ldots, \hat{\theta}_{\ell-1})$ be the exposed secret elements of the targeted \WOTSP\ instance, corresponding to message chunks $(\hat{m}_0, \hat{m}_1, \ldots, \hat{m}_{\ell-1})$.
Following the approach by Castelnovi et al.~\cite{castelnoviGraftingTreesFault2018}, the algorithm to find a suitable \XMSS\ tree works as follows:
\begin{enumerate}
    \item Create a new \XMSS\ tree with public key $\tilde{\pk}{}^X$ from a secret $\hatSKseed$ drawn uniformly at random.
    \item Split $\tilde{\pk}{}^X$ and its \WOTSP\ checksum into chunks $(\tilde{r}_0, \ldots, \tilde{r}_{\ell-1})$ of size $w$ bits.
    \item If $\tilde{r}_i \geq \hat{m}_i$ for all $0 \leq i < \ell$, return the grafted $\hatSKseed$. Repeat from step 1 otherwise.
\end{enumerate}
\emph{Complexity}:
The hash complexity of tree grafting may impose significant computational cost, scaling with the number of signatures for a compromised \WOTSP\ instance and the number of compromised \WOTSP\ secret values.
In our evaluation (\cref{sec:evaluation}), the observed grafting complexity ranges from $2^{27.98}$ hashes (SHAKE-128f deterministic) to $2^{130.46}$ hashes (SHAKE-128s deterministic) for parameter sets where faults were collected, with successful forgeries achieved for complexities up to $2^{32.95}$ (SHA2-128s randomized).
The total attack complexity, combining grafting and seeking, determines the practical feasibility of the forgery.
In \cref{sec:exact_grafting_complexity}, we describe how to determine the actual complexity of attacking specific faulted instances.

\subsection{Path Seeking}\label{sec:path_seeking}
The randomization value $R$, together with the message $m$ to be signed, determines the path taken through the hypertree.
This randomization value is determined by the signing party, and can therefore be chosen by the attacker.
For the universal forgery, they have to find a value $R'$ such that the signing visits the compromised \WOTSP\ instance and the grafted \XMSS\ tree.
Finding such a value $R'$ boils down to searching an $n$ byte number such that the $h-h'l^*$ most significant bits are equal to the index of the grafted subtree.\\
\emph{Complexity}:
Path seeking yields cost for each forged signature.
Each attempt at finding a valid $R'$ requires a single hash function call, and the attacker on average needs $2^{h-h'l^*}$ hash function calls to find a suitable $R'$.

\section{An Exact Solution For Grafting Complexity}\label{sec:exact_grafting_complexity}

Despite previous analysis on the theoretical grafting complexity for a given \SLHDSA\ instance, the literature currently lacks a method for estimating the grafting complexity for a concrete compromised \WOTSP\ key pair after the collection of faulty signatures.
While Gen\^et provides a sound estimate for the average grafting complexity for uniformly distributed \WOTSP\ secrets~\cite{genetProtectingSPHINCSFault2023}, in practice, an attacker has to choose from a concrete set of \WOTSP\ instances to graft a tree for.
However, for concrete \WOTSP\ instances, there can be a wide variance in grafting complexity even for the same number of observed collisions.
Our analysis closes this gap, allowing an attacker not only to estimate the runtime for the offline grafting phase, but also to rank compromised \WOTSP\ key pairs based on the expected forgery runtime.
We introduce a novel combinatorial technique to precisely determine the grafting success probability for a given set of exposed \WOTSP\ secret values, enabling attackers an improved estimation of the total attack complexity.
This approach extends Gen\^et's work by providing exact complexity analysis for concrete instances rather than average-case estimates.

\subsection{Notation}\label{subsec:notation}

Let $w \in \mathbb{Z}_+$ be the Winternitz parameter.
For a \WOTSP\ signature $\sig$ of a message $m$, we denote the message chain indices in $\sig$ corresponding to message block $i \in \{0,1,\dots,\ell_1-1\}$ by $m_i \in \{0, \dots, w-1 \}$.
Similarly, denote the checksum chain indices corresponding to the checksum $c$ of $m$ by $c_i \in \{0, \dots, w-1 \}$ for $i \in \{0, \dots, \ell_2-1\}$.
We denote a full \WOTSP\ chain by $\theta = (m_0, m_1, \dots, m_{\ell_1-1}, c_0, c_1, \dots, c_{\ell_2-1})$.

For chain $i$, we call the preimages $0 \leq \tilde{\theta}_i < \theta_i$ of $\theta_i$ the \emph{\WOTSP\ secret values}.
If the attacker obtains two or more \WOTSP\ signatures for distinct messages such that at least one secret value is exposed, we call this instance \emph{compromised}.

\subsection{A Combinatorial Solution For Grafting Complexity}

There is a wide variance in the resulting grafting complexity even for the same number of collisions in a \WOTSP{} instance.
Therefore, we now derive the expected runtime for forging a signature using a specific key pair.
To exploit a compromised \WOTSP\ instance, the attacker combines the secret values exposed by signatures from that instance, collecting the minimal chain secret values for each chain.
They then use the exposed chain secrets to graft and sign an \XMSS\ tree at the next layer.
However, they have to find a tree matching the exposed chain secrets.
Since the sampled \XMSS\ trees are distributed uniformly at random, the attacker can now compute the probability that a random message can be signed given the exposed chain secrets, a proxy for the post-processing \emph{grafting complexity}.
As a \WOTSP\ signature is only valid if message and checksum match, they have to find the number of valid message-checksum combinations.
Enumerating all messages and filtering for the ones reachable with the available checksums is infeasible, because there are $w^{\ell_1}$ possible messages, ranging from $2^{128}$ to $2^{256}$ depending on the parameter set.
However, since $\ell_2=3$ for all parameter sets, they can enumerate the reachable checksums and compute the number of valid messages for those checksums.
We first formulate this challenge as a combinatorial problem and then introduce an implementation based on dynamic programming.
For that, we denote the integer representation of a checksum~$c$ regarding a base $w$ by $\toInt(c, w) = \sum_{i=0}^{\ell_2} c_i \cdot w^{i}$.

\textbf{Chain capacities:}
We define the per-component capacities of message chain $i=0, \dots, \ell_1-1$ as $k_i = w - 1 - m_i$.
These capacities describe the number of times each message chain can be iterated until the chain's end at step $w-1$ is reached.

\textbf{Enumeration of checksums:}
Let $\mathcal{T}$ be the set of checksums computable by continuing checksum chains until the chain's end or the maximum checksum $\ell_1 \cdot (w-1)$ is reached.
More formally, $\mathcal{T} \subseteq \{ 0, 1, \dots, w-1 \}^{\ell_2}$ consist of the tuples $\tau = (\tau_0, \dots, \tau_{\ell_2-1})$ satisfying for all $0 \leq i \leq \ell_2-1$:
$$c_i \leq \tau_i < w~ \land~ \toInt(\tau, w) \leq \ell_1 \cdot (w-1).$$

\textbf{Counting signable compositions:}
For each valid tuple $\tau \in \mathcal{T}$, define the capacity to be distributed across chains as $\kappa(\tau) = \toInt(c, w) - \toInt(\tau, w)$.
We now want to determine the number of ways the capacity can be distributed under the constraints $\mathcal{K} = \{ k_i \mid 0 \leq i < \ell_1 \}$, i.e., the number of compositions $x = (x_0, x_1, \dots, x_{\ell_1})$ satisfying $\sum_{i=0}^{\ell_1} x_i = \kappa(\tau, \mathcal{K}) \text{ such that } 0 \leq x_i \leq k_i$.
Generating solutions to this system is known as the \emph{second-order restricted weak integer composition}~\cite{pageGeneralizedAlgorithmRestricted2013} problem and, in general, there are exponentially many solutions to this system.
However, we are only interested in the number of compositions $N(\kappa(\tau, \mathcal{K}))$ satisfying this system, which can be computed in time $\mathcal{O}(\tau \cdot \ell_1)$ and memory $\mathcal{O}(\tau)$ by the dynamic programming based algorithm shown in \cref{alg:compositions}.

\textbf{Putting it all together:}
Using $N(\kappa(\tau, \mathcal{K}))$ for all $\tau \in \mathcal{T}$, we can compute the probability that a random \XMSS\ tree can be signed with a given compromised \WOTSP\ instance by taking the fraction of signable messages over the number of all messages:
$\mathbb{P}\left(\text{Signable}\right) = \frac{\sum_{\tau \in \mathcal{T}} N(\kappa(\tau, \mathcal{K}))}{w^{\ell_1}}$.
Modeling tree grafting as a Bernoulli process with success probability $\mathbb{P}\left(\text{Signable}\right)$, we can, on average, expect the tree grafting to succeed after $\mathbb{C}(\text{Grafting}) = \frac{1}{\mathbb{P}\left(\text{Signable}\right)}$ attempts.
Multiplying $\mathbb{C}(\text{Grafting})$ with the number of hash operations required to compute an \XMSS\ tree yields the grafting hash complexity.

\section{Evaluation of SLasH-DSA}
\label{sec:evaluation}

In this section, we evaluate the practical feasibility and effectiveness of our Rowhammer-based attack against \SLHDSA\ signature schemes.
We leverage the \tech{Swage} framework introduced in \cref{cha:swage} to orchestrate the end-to-end attack: based on our fault analysis in \cref{sec:openssl_fault_analysis}, we implemented an \SLHDSA-specific \tech{Orchestrator} module for \tech{Swage}, and use \tech{Swage}'s core components to perform the attack: \tech{Swage} handles DRAM address reverse-engineering, hammering pattern execution, page injection to co-locate victim data with aggressor rows, and signature collection during the online phase.
Taking advantage of \tech{Swage}'s modular architecture, this allows us to focus on the target-specific parts of the attack, relying on established attack primitives implemented in the \tech{Swage} core.
We demonstrate the ability to successfully compromise the deterministic and randomized variants of \SLHDSA\ across different parameter sets.
Our evaluation quantifies the computational complexity of the grafting and seeking phases, and analyzes the trade-offs between different attack strategies.
The results show that our attack can achieve practical signature forgery within reasonable time bounds for some \SLHDSA\ parameter configurations, taking only a few minutes to an hour on commodity hardware, while showing theoretical attack complexities for most parameter sets.

\subsection{Experimental Setup}

We conducted our experiments on a system with an Intel i5-6400 CPU and a single G.Skill AEGIS 16 GB DDR4-2133 DIMM, running Ubuntu 20.04.6 with Linux kernel version 6.8.12-generic with default settings.
We evaluated the \SLHDSA\ implementation in OpenSSL version 3.5.1, released on July 1st, 2025.
The signing process was implemented via a small signing server accepting signing requests via standard input and returning signatures to standard output.
The signing server is configurable with all standardized \SLHDSA\ parameter sets with both random and deterministic signing.
Both the signing server and OpenSSL were built with gcc version 11.4.0 with all \texttt{-O3} optimizations enabled.
The post-processing is performed on an Intel Xeon Gold 5415+ with 128~GB RAM, running Ubuntu 20.04.6 with Linux kernel version 5.11.0-generic with default settings.

While the experimental setup we chose demonstrates the feasibility of the attack, newer hardware incorporates novel Rowhammer defenses, and newer software versions may include additional countermeasures.
Future work should investigate the attack's effectiveness on more recent and varying system configurations.

\subsection{Attack Against Deterministic \SLHDSA}

\newcolumntype{R}[2]{%
    >{\adjustbox{angle=#1,lap=\width-(#2)}\bgroup}%
    l%
    <{\egroup}%
}
\newcommand*\rot{\multicolumn{1}{R{45}{1em}}} %
\newcommand{\expthead}{\rot{Signatures} & \rot{Faulted} & \rot{Layer} & \rot{Grafting} & \rot{Seeking} & \rot{Time (s)}}

\newcommand{\hashfncol}[1]{\parbox[c]{2mm}{\multirow{6}{*}{\rotatebox[origin=c]{90}{#1}}}}

\begin{table*}[t]
\centering
\caption{
  Best total attack complexity for one-shot forgery after (a) one hour of hammering against deterministic signing and (b) eight hours of hammering against randomized signing with all parameter sets.
  We consider the attack complexity to be the sum of grafting and seeking for a one-shot forgery attack, i.e., signing one attacker-selected message.
  Grafting and Seeking columns show hash complexity (cf. \cref{sec:exact_grafting_complexity}), time is in seconds.
  Empty cells (--) indicate parameter sets where insufficient faulted signatures were collected to carry out an attack within a reasonable amount of time.
  }

\subcaptionbox{\label{tab:det_sign_compl}}{
  \centering
  \begin{tabular}{
  ll@{\hspace*{0.5em}}  %
  ll@{\hspace*{0.5em}}  %
  *{4}{c}  %
  }
  \toprule
  & & \expthead \\ \midrule%
  \hashfncol{SHA2}
  & 128s & $187$ & 1 & $3$ & $2^{93.37}$ & $2^{36}$ & -- \\
  & 128f & $1917$ & 1 & $6$ & $2^{30.66}$ & $2^{48}$ & -- \\
  \cmidrule{2-8}%
  & 192s & $105$ & 2 & $5$ & $2^{52.32}$ & $2^{18}$ & -- \\
  & 192f & $567$ & 7 & $12$ & $2^{42.09}$ & $2^{30}$ & -- \\
  \cmidrule{2-8}%
  & 256s & $115$ & 3 & $6$ & $2^{53.85}$ & $2^{16}$ & -- \\
  & 256f & $275$ & 3 & $4$ & $2^{50.18}$ & $2^{52}$ & -- \\
  \midrule
  \hashfncol{SHAKE} 
  & 128s & $163$ & 2 & $4$ & $2^{130.46}$ & $2^{27}$ & -- \\
  & 128f & $1683$ & 3 & $17$ & $2^{27.98}$ & $2^{15}$ & 151 \\
  \cmidrule{2-8}%
  & 192s & $95$ & 1 & -- & -- & -- & -- \\
  & 192f & $517$ & 10 & $21$ & $2^{36.56}$ & $2^{3}$ & -- \\
  \cmidrule{2-8}%
  & 256s & $105$ & 5 & $3$ & $2^{53.20}$ & $2^{40}$ & -- \\
  & 256f & $253$ & 10 & $15$ & $2^{44.53}$ & $2^{8}$ & -- \\
  \bottomrule%
  \end{tabular}%
}\hfill%
\subcaptionbox{\label{tab:rnd_sign_compl}}{
  \begin{tabular}{
  ll@{\hspace*{1em}}  %
  ll@{\hspace*{1em}} %
  *{4}{c}  %
  }
  \toprule%
  & & \expthead \\ \midrule%
  \hashfncol{SHA2}
  & 128s & $1503$ & 37 & $6$ & $2^{32.95}$ & $2^{9}$ & 3781 \\
  & 128f & $7659$ & 1 & -- & -- & -- & -- \\
  \cmidrule{2-8}%
  & 192s & $843$ & 50 & $6$ & $2^{41.91}$ & $2^{9}$ & -- \\
  & 192f & $4617$ & 6 & $19$ & $2^{40.84}$ & $2^{9}$ & -- \\
  \cmidrule{2-8}%
  & 256s & $949$ & 102 & $7$ & $2^{51.59}$ & $2^{8}$ & -- \\
  & 256f & $2279$ & 26 & $15$ & $2^{52.65}$ & $2^{8}$ & -- \\
  \midrule
  \hashfncol{SHAKE}
  & 128s & $1323$ & 3 & -- & -- & -- & -- \\
  & 128f & $13479$ & 33 & $18$ & $2^{85.85}$ & $2^{12}$ & -- \\
  \cmidrule{2-8}%
  & 192s & $761$ & 360 & $6$ & $2^{36.42}$ & $2^{9}$ & -- \\
  & 192f & $4137$ & 209 & $21$ & $2^{31.29}$ & $2^{3}$ & 1417 \\
  \cmidrule{2-8}%
  & 256s & $869$ & 26 & $7$ & $2^{52.77}$ & $2^{8}$ & -- \\
  & 256f & $2035$ & 34 & $15$ & $2^{52.31}$ & $2^{8}$ & -- \\
  \bottomrule%
  \end{tabular}%
}%
\end{table*}

At the start of the signing procedure, the \SLHDSA\ signing algorithm determines the \FORS\ instances for signing the message and the path taken through the hypertree.
The \FORS\ instances and path through the hypertree depend on the message to be signed, the \PKseed\ and \PKroot, and an optional random value $R$.
The randomization of the signing process is optional, and is set to $\PKseed$ when deterministic mode is selected.
We start the experiments with an attack against the deterministic variant of \SLHDSA.
\cref{tab:det_sign_compl} shows the attack complexity after one hour of hammering. %
The table shows the number of signatures collected and the number of observed faults (i.e., signatures not matching the expected output from the signing server).
For the most exposed \WOTSP\ instance, it reports the layer this instance was encountered, the grafting and seeking complexities in number of hashes and, if applicable, the post-processing time for a successful forgery.
Our results highlight the increased susceptibility of deterministic \SLHDSA, as deterministic signing maximizes the number of possible collisions due to the reuse of \FORS, \XMSS, and \WOTSP\ instances throughout the signing procedure.

\subsection{Attack Against Randomized \SLHDSA}

In contrast to the deterministic variants, the randomized \SLHDSA\ signature scheme incorporates an additional random value $R$, which further diversifies the signing process even for signing the same message repeatedly.
This randomization significantly complicates the attack methodology, as it disrupts the reuse of specific cryptographic instances that we exploit in the deterministic mode.
In this subsection, we outline our approach to adapting collision-based strategies to randomized signing.
We analyze the impact of the extra randomness on both the grafting and path seeking phases, and discuss the increased computational challenges introduced by randomization.
Our experiments provide insights into the effectiveness and limitations of the attack under these conditions.

For deterministic signing, a fixed message always results in the same path through the hypertree, whereas for randomized signing, the path is determined using $R$.
Due to the randomized path, it is unlikely that \XMSS\ trees at low hypertree levels will be reused, and we cannot expect to find colliding \WOTSP\ instances for our attack on these levels.
This is especially true for the ``small'' parameter sets of the scheme: For example, in the 256s parameter set with \XMSS\ tree height $h'=8$, there are $256$ \WOTSP\ instances in the topmost layer $16$, and $256^{15-l}$ at layer $0 < l < 16$.
The ``fast'' variants, however, stand out with small \XMSS\ trees and more hypertree layers.
For example, the 192f parameter set has \XMSS\ tree height $h'=3$, resulting in $8$ \WOTSP\ instances in the topmost layer $21$, and $8^{20-l}$ at layer $0 < l < 21$.
This makes \WOTSP\ collisions on layers other than the topmost layer feasible, increasing the attack surface of our attack against randomized signing.
\cref{tab:rnd_sign_compl} shows the attack cost for a one-shot forgery attack, i.e., forging a valid signature for one attacker-chosen message, after eight hours of hammering the randomized variants of \SLHDSA\ for all parameter sets.
While the number of faulted signatures serve as a first proxy for the success of an attack, the fault characteristic is important for the attack:
for example, faulting the hash function's internal constants will not produce useful faults for the attacker.
We presume that the increased fault rate for some parameter sets, such as SHAKE-192s, where we observe a significant number of faulted signatures not corresponding to improved attack complexities, are due to this effect.
We hypothesize however, that longer hammering time in general improves attack capabilities.

The experiments against both deterministic and randomized signing were conducted while applying performance degradation as described in \cref{sec:performance_degradation}.
The experiments show that the end-to-end attack is feasible after only a few hours of hammering for some parameter sets, resulting in a post-processing time for both deterministic and randomized signing of only a few minutes.
The grafting step complexity is calculated by applying our novel, exact solution method from \cref{sec:exact_grafting_complexity}.
Weighing grafting vs. path seeking is a promising optimization for multi-shot signing, as the cost for path seeking must be spent on each message, while grafting is only needed once.

For unsuccessful forgeries in \cref{tab:det_sign_compl,tab:rnd_sign_compl}, we identify two bottlenecks: for the SHAKE-192s (deterministic), SHA2-128f, and SHAKE-128s (randomized) parameter sets, insufficient faulted signatures were collected during the hammering campaigns, resulting in no compromised \WOTSP{} instances.
Besides, for parameter sets with compromised \WOTSP{} instances but high grafting complexities, the offline post-processing phase is computationally infeasible on commodity hardware (e.g., $2^{85.85}$ hashes for SHAKE-128f randomized).
However, using a more vulnerable DIMM with higher bit flip rates could improve fault collection during the online phase, potentially making attacks feasible for additional parameter sets.
This is in contrast to Gen\^et's hardware-based approach~\cite{genetProtectingSPHINCSFault2023}, where glitch injection enables attacks in all five trials of his practical attack against the SHAKE-256s parameter set with 1024 collected signatures per trial, the attack being restricted solely by the offline post-processing.

\section{Related Work}
\label{sec:related_work}
In this section, we discuss prior work relevant to our study, including grafting tree attacks and fault analysis techniques, Rowhammer-based attacks that target \PQC{} schemes, and existing countermeasures against Rowhammer vulnerabilities.

\subsection{Grafting Attacks and Countermeasures}
Castelnovi et al.~\cite{castelnoviGraftingTreesFault2018} were the first to present a theoretical analysis of the \emph{grafting tree attack} including computational feasibility.
Gen\^et et al. implemented the grafting tree attack against SPHINCS~\cite{genetPracticalFaultInjection2018}.
Gen\^et later presented a theoretical analysis of the grafting tree attack against \SPHINCSP\ along with a practical attack implementation~\cite{genetProtectingSPHINCSFault2023}.
The fault injection method in those works is a hardware-supported clock glitching attack focused on embedded devices.

Additionally, Gen\^et \cite{genetProtectingSPHINCSFault2023} examined the \emph{layer caching} and \emph{branch caching} countermeasures against grafting tree attacks.
The layer caching countermeasure involves pre-computing upper layers of the \SPHINCSP\ hypertree, storing them alongside the public key.
This reduces the number of attackable hash function calls, forcing the attacker to find collisions in the lower layers, decreasing collision probability and increasing path seeking complexity for successful faults.
This countermeasure shows varying degrees of effectiveness, depending on the number of cached layers.
With the \emph{branch caching} countermeasure, on the other hand, the signing party stores all encountered \WOTSP\ signatures and public keys in a fixed-size cache.
However, it was shown that this countermeasure is ineffective in protecting \SLHDSA, as cache misses and therefore recomputations of \WOTSP\ instances are inevitable with reasonably sized caches.
To summarize, the best current solution to protect the scheme is \emph{redundancy}, where (parts of) the signature is verified by duplicate computation \cite{castelnoviGraftingTreesFault2018,amietFPGAbasedSPHINCSImplementations2020}.

\subsection{Rowhammer Attacks against Post-Quantum Schemes}
Rowhammer has also shown effective as a \emph{fault injection method against other post-quantum schemes}.
In PQHammer~\cite{amerPQHammerEndtoEndKey2025}, Amer et al. describe end-to-end key recovery attacks based on Rowhammer against the post-quantum key encapsulation mechanisms (KEMs) BIKE and CRYSTALS-Kyber (ML-KEM), and against the lattice-based signature scheme CRYSTALS-Dilithium (ML-DSA).
In Crowhammer~\cite{haidarCrowhammerFullKey2025}, Haidar et al. show how to attack the post-quantum signature scheme Falcon to perform a Rowhammer-based key recovery attack.
This line of research highlights the need to include Rowhammer attacks against post-quantum cryptography in the threat models.
A starting point for this are tools like HammerTime~\cite{tatarDefeatingSoftwareMitigations2018}, enabling profiling and simulation of Rowhammer attacks.
While the profiling results indicate how many hammering experiments are required to flip a bit in a given offset range within a page, HammerTime does not cover all necessary steps for an end-to-end Rowhammer attack.

\subsection{Fault Analysis}
Orthogonal to exploiting memory fault vulnerabilities, another line of work \emph{analyzes the vulnerability of schemes to fault attacks.}
Achilles~\cite{liangAchillesFormalFramework2025} systematically analyzes signature schemes by separating the fault model from the algorithm under analysis:
A generalized signature scheme utilizes public and secret parameters combined with signing oracles that generate valid and faulty signatures.
A fault model then specifies where to inject the faults, and in a post-processing, the secret key is recovered.
The authors showcase the efficacy of Achilles in the key recovery domain with six different signature schemes including the post-quantum scheme CRYSTALS-Dilithium (ML-DSA).
However, transitioning from key recovery to universal forgery of \SLHDSA\ is not straightforward as there is no obvious reduction from their attacker model for deriving secret values to forging valid signatures.
Rainbow~\cite{loossIntegratingFaultInjection2022} is another automated fault analysis tool that emulates the target program with the Unicorn CPU emulator~\cite{unicornUnicornUltimateCPU2015}.
Rainbow allows to specify fault injection parameters for automated fault injection against a target program.

\subsection{Rowhammer Countermeasures}
There has been a variety of \emph{Rowhammer countermeasures} implemented by manufacturers and proposed in scientific work.
\TRR{}, a DDR4 vendor-specific mechanism, tracks memory accesses and issues refresh commands on unusual access patterns, preventing simple Rowhammer attacks by preemptively restoring charge before bit flips can be induced.
However, \TRR{} is ineffective against advanced hammering patterns~\cite{frigoTRRespassExploitingMany2020,jattkeBlacksmithScalableRowhammering2022}.
Hardware-supported \ECC{} can detect and correct bit flips~\cite{seabornProjectZeroExploiting2015, grussAnotherFlipWall2018}, but increases \DRAM{} cost and may be ineffective against certain Rowhammer attacks~\cite{cojocarExploitingCorrectingCodes2019,kamadanECCfailMountingRowhammer2025} or introduce new attack vectors such as timing-based attack~\cite{kwongRAMBleedReadingBits2020}.
Even DDR5, while introducing new countermeasures \cite{DBLP:journals/corr/abs-2406-19094,jedec2024ddr5c}, has been found to be susceptible to Rowhammer attacks \cite{jattkeZenHammerRowhammerAttacks2024a,jattkeMcSeeEvaluatingAdvanced2025}.

OS-level mitigations like \emph{guard rows} around sensitive memory pages~\cite{veenGuardIONPracticalMitigation2018,konothZebRAMComprehensiveCompatible2018} prevent some Rowhammer attacks but introduce high memory and performance costs.
They can furthermore be bypassed by attacks using non-adjacent aggressor rows~\cite{jattkeBlacksmithScalableRowhammering2022,koglerHalfDoubleHammeringNext2022}.
However, all of these countermeasures have individual limitations, are not universally applicable, or are not effective against all Rowhammer attacks.

\section{Future Work}
\label{sec:future_work}

In this paper, we demonstrated the feasibility of a Rowhammer attack against \SLHDSA\ as an exemplary post-quantum signature scheme.
We implement the \tech{Swage} framework, enabling easy implementation of practical Rowhammer attacks.
One remaining aspect requiring substantial manual effort with \tech{Swage} is analyzing the binary and scheme to find memory locations susceptible to Rowhammer fault attacks.
This could be improved by integrating \tech{Swage} more closely with tools like Achilles or Rainbow (c.f. \Cref{sec:related_work}).

As shown in \Cref{tab:det_sign_compl} and \Cref{tab:rnd_sign_compl}, for some parameter sets, tree grafting and path seeking 
require significant CPU resources assuming a limited amount of hammering.
Since both tree grafting and path seeking are heavily dependent on the hashing performance, a substantial speedup is to be expected with GPU acceleration.
In 2025, Wang et al. presented CUSPX~\cite{wangCUSPXEfficientGPU2025}, a GPU-accelerated implementation of SPHINCS\textsuperscript{+}.
Preliminary benchmarks using their hash function implementations show a substantial increase in hash performance, achieving $2^{37.2}$ SHA-256 operations per second on an NVIDIA RTX 4090.
This marks a $2^{14}$-fold increase compared to our evaluation system.
A GPU-accelerated implementation of the grafting and path seeking algorithms should speed up the attack significantly, potentially allowing to break some of the remaining parameters sets.

Moreover, for the attack presented in this paper, we assume that the attacker is core co-located with the victim for the page injection.
In Rubicon~\cite{bolcskeiRubiconPreciseMicroarchitectural2025}, Bölcskei et al. demonstrate that page injection attacks also work across CPUs, widening the threat model to be considered for Rowhammer attacks and potentially making a wider range of systems vulnerable.

\section{Conclusion}
\label{cha:conclusion}

We presented SLasH-DSA, an end-to-end Rowhammer attack against \SLHDSA.
We demonstrated this attack against OpenSSL to be feasible and effective in practice, allowing an attacker to conduct a universal forgery attack against all NIST security levels of \SLHDSA\ by flipping bits in the signature generation process.
In addition, we provided a novel complexity analysis for concrete faulted signatures that allows steering the post processing along the path of lowest complexity, speeding up the attack.
This is, to the best of our knowledge, the first practical demonstration of a Rowhammer attack against \SLHDSA\ on a real-world system.
To conduct the attack, we introduced \tech{Swage}, a comprehensive open-source framework for performing Rowhammer attacks on real-world systems.
\tech{Swage} provides a modular interface for Rowhammer attacks, allowing users to focus on the attack logic rather than the underlying hardware and operating system details.
To our knowledge, \tech{Swage} is the first fully open-source end-to-end framework for Rowhammer attacks, providing a complete and extensible solution for attacks on real-world systems.

\ifanon
\else
\section{Acknowledgements}
This work has been supported by BMFTR through the AnoMed project and by the BMBF project SASVI.

\fi

\bibliographystyle{plain}
\bibliography{ref.bib}

\section{Appendix}
\label{sec:appendix}

\subsection{Number of Weak Compositions}

The following DP-based algorithm computes the number of weak compositions $\sum_{i=0}^{\ell_1-1} x_i$ for a target sum $\tau$ respecting upper bounds $k_i$ for $0 \leq i \leq k_{\ell_1-1}$:

\begin{algorithm}
\caption{\textsc{Weak-Comp}$(\tau,\;k_0,\dots,k_{\ell_1-1})$}\label{alg:compositions}
\KwIn{Target sum $\tau$, upper bounds $k_0,\dots,k_{\ell_1-1}$}
\KwOut{Number of weak compositions $x=(x_0, \dots, x_{\ell_1-1}$) of $\tau$ respecting $0 \leq x_i \leq k_i$}
\DontPrintSemicolon
\SetAlgoLined
\SetAlgoNoEnd

\BlankLine
\textbf{Initialize}\;
\Indp
\For{$t \gets 0$ \KwTo $\tau$}{
    $dp[t] \gets 0$\;
}
$dp[0] \gets 1$\tcp*{one way to compose 0}
\Indm

\BlankLine
\For{$i \gets 0$ \KwTo $\ell_1-1$}{
    $\mathrm{prefix} \gets 0$\;
    \For{$t \gets 0$ \KwTo $\tau$}{
        $\mathrm{prefix} \gets \mathrm{prefix} + \mathrm{dp}[t]$\;
        \lIf{$t > k_i$}{$\mathrm{prefix} \gets \mathrm{prefix} - \mathrm{dp}[t - (k_i + 1)]$}
        $\mathrm{new}[t] \gets \mathrm{prefix}$\;
    }
    \For{$t \gets 0$ \KwTo $\tau$}{
        $\mathrm{dp}[t] \gets \mathrm{new}[t]$\;
    }
}
\Return $dp[\tau]$\;
\end{algorithm}

\end{document}